\documentclass[fleqn,usenatbib]{mnras}
\usepackage{graphicx}   
\usepackage{amsmath}    
\usepackage{amssymb}    
\usepackage{natbib}
\usepackage{color}
\usepackage{txfonts}
\definecolor{myred}{rgb}{0.6,0.1,0.1}
\definecolor{mybrown}{rgb}{0.4,0.18,0.06}

\newcommand\corr[1]{{\color{myred}\ #1}}

\title{Planet-planet scattering in presence of a companion star}
\author[]{
Francesco Marzari$^{1}$\thanks{E-mail: francesco.marzari@pd.infn.it}, 
Makiko Nagasawa$^{2}$\thanks{E-mail: nagasawa\_makiko@med.kurume-u.ac.jp},
Krzyszof~Go\'zdziewski$^{3}$\thanks{e-mail: k.gozdziewski@umk.pl}
\\
$^{1}$Department of Physics and Astronomy, University of Padova, Via Marzolo 8, 35131 Padova, Italy\\
$^{2}$ Kurume University, School of Medicine, Department of Physics, 67 Asahi--machi, Kurume--city, Fukuoca, 830--0011, Japan. \\
$^{3}$Institute of Astronomy, Faculty of Physics, Astronomy and Informatics, N. Copernicus Univ., Grudziadzka 5, 87-100 Toru\'n, Poland. \\
}
\date{\today{}}

\begin{document}

\pagerange{\pageref{firstpage}--\pageref{lastpage}}

\maketitle
\begin{abstract}

Planet-Planet (P--P) scattering is a leading dynamical mechanism invoked to explain the present orbital distribution of exoplanets. Many stars belong to binary systems, therefore it is important to understand how this mechanism works in presence of a companion star. We focus on systems of three planets orbiting the primary star and estimate the timescale for instability finding that it scales with the keplerian period for systems that have the same ratio between  inner planet and binary semi--major axes. An empirical formula is also derived from simulations to estimate how the the binary eccentricity affects the extent of the stability region.

The presence of the secondary star affects the P--P scattering outcomes causing a broadening of the final distribution in semi--major axis of the inner planet as some of the orbital energy of the planets is absorbed by the companion star. Repeated approaches to the secondary star causes also a significant reduction in the frequency of surviving two--planet systems in particular for larger values of the inner planet semi--major axis. The formation of Kozai states with the companion star increases the number of planets which may be tidally circularized.  
To predict the possible final distribution of planets in binaries we have performed a large number of simulations where the initial semi--major axis of the inner planets is chosen randomly.  For small values of the binary semi--major axis, the higher frequency of collision alter the final planet orbital distributions which, however, beyond 50 au appear to be scalable to wider binary separations. 

\end{abstract}

\begin{keywords}
Planetary systems -- planets and satellites: general -- planets and satellites: dynamical evolution and stability 
\end{keywords}

\label{firstpage}
\section{Introduction}

Planet--Planet scattering (hereinafter P--P scattering) is believed to be one of the most important mechanism in sculpting planetary systems after the phase of planet formation (\cite{weimar1996,rasioford1996,lin1997} see also \cite{davies2014} for a review). The chaotic evolution caused by the repeated planetary encounters lead to the formation of highly eccentric and inclined planets which can explain the currently observed exoplanet orbital distribution.
A multiple gas giant system, after a period of P-P scattering, may also end up in a configuration with a Hot/Warm Jupiter. An inward scattered planet may settle down in a close low eccentricity orbit after  circularization to the pericenter by tidal dissipation of energy \citep{naga08,beauge2012}. In a recent paper, \cite{marzanaga2019} have shown that after the inclusion in the numerical model of the effects of planet collisions, tidal interactions with the central star and general relativity, populations of initially three unstable planets can explain the observed eccentricity and semi--major axis distribution of giant planets close to their star i.e. within about 2 au. It has also been suggested that P--P scattering may have occurred in our solar system where an additional planet(s), now escaped, would have triggered a period of  instability which could explain the Main Belt excitation \citep{wetherill1992, deienno2018}, {the eccentricity excitation of the planets and the capture of Jupiter Trojans and its satellites (see \cite{nesvorny2018} for a review of the subject) }.

Due to the relevance of P--P scattering in the history of planetary systems, it is important to understand how the presence of a binary companion may affect its occurrence and final outcome. It is known that a considerable fraction, ranging from 40\% to 45\% of sun--like star in the solar neighborhood belong to binary or multiple systems \citep{dun1991,raghavan2010}.  Exoplanets have also been detected in binary systems and some statistical studies \citep{desidera2007,mug2009,roell2012} have estimated that about 10\% to 15\% of binaries may host planets. An updated list of planets in multiple stellar systems can be found  in the  "Catalogue of Exoplanets in Binary and Multiple Star Systems" of \cite{schwarz2016} {($www.univie.ac.at/adg/schwarz/bincat\_binary\_star.html$)}.  Given that planet formation in binaries appears to have a rate which is comparable to that around single stars, an important question is to understand how the dynamics and evolution of these planetary systems is influenced by the gravity of the companion star. Here we focus on the P--P scattering mechanism involving three initial planets and we try to explore the possible effects that a companion star may have on their evolution.  In a previous paper, \cite{marza2005ApJ} found a higher percentage of ejection of one or two planets out of the system due to the perturbations of the secondary star possibly leading to less populated planetary systems in binaries compared to single stars. They performed the modeling in a system with a binary semi--major axis of 50 au and varied the binary eccentricity to test the dependence of the P--P scattering outcomes on this parameter.  

In this paper, we perform P--P scattering simulations in binaries with a more refined code which includes the effects of tides and general relativity and consider a wider range of initial conditions for the initial three--planet system. In addition, we also look for scaling relationships which may allow to generalize the results for a single configuration to a wider range in the parameter space. This is an important step to reduce the large number of free parameters which plague this particular dynamical problem. 

In Sect. 2 we describe the initial configuration of the planetary system leading to P--P scattering.  In Sect. 3 we introduce a scaling of the dynamics with $R$, the ratio between the inner planet semi--major axis and that of the binary. This scaling works both for the instability times, which also scale with the Keplerian period of the inner planet, and for the long--term stability properties of the system computed with MEGNO (Sect. 4).  In Sect. 5 we illustrate how the stability regions depend on the binary eccentricity, once fixed $R$, and derive a semi--empirical relation between the binary eccentricity and the limiting value of $R$ for the presence of  stable regions.
In Section 6 we briefly describe the numerical model for P--P scattering which includes tides, collisions and general relativity. In Sect. 7 we summarise the types of possible outcomes of the chaotic phase and in Sect.~8  we compare the final distribution of P--P scattering in a reference case with and without the presence of a binary companion to outline the differences.  In Sect. 9 statistical results derived from initial random choices of the planets orbital elements are presented.  Sect. 10 is devoted to a more realistic modeling where the initial semi--major axis of the inner planet is  randomly sampled before the close encounter phase. In this way we try to predict what the observed distribution of planets in binaries could be after de--biasing the observations.   Finally, in Sect. 11 we discuss our results.

\section{The dynamical configuration}

\begin{figure}
\centering
\includegraphics[width=0.9\columnwidth]{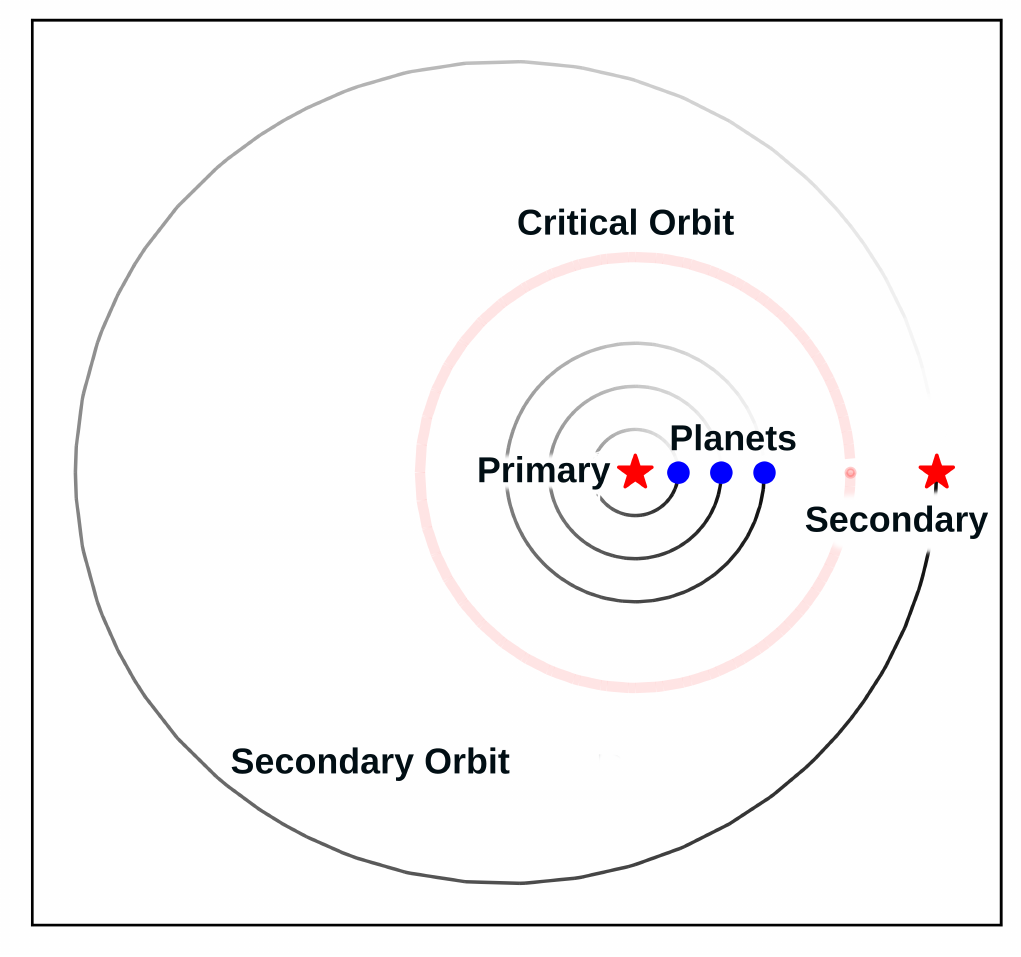}
\caption{
Geometry of the system in the astrocentric reference frame centered at the primary star. The binary eccentricity is $e_B=0.3$. The light-red curve illustrates the critical orbit for stability derived from the  semi--empirical formula of 
Wisdom and Holman (1999).
}
\label{fig:geometry}
\end{figure}

In our models we consider a scenario with  three planets orbiting the primary star of a binary star system. The planets are in an unstable configuration due to the mutual gravitational perturbations and those of the  companion star. The instability may be triggered just after the dissipation of the protoplanetary disk in which the planets were embedded as the friction with the gas, able to damp the eccentricities of the planets, become ineffective. In alternative, the planets may start to have close encounters at later times, even billions of years after the dissipation of the gaseous disk,  because the system is intrinsically chaotic. 
The scenario we are exploring is illustrated in 
Fig.\ref{fig:geometry} showing the initial configuration of the planets in S--type orbits (non-circumbinary planets) around one of the stars. 
We will not investigate the way in which these kind of planetary systems formed and possibly migrated by interaction with the disk. The binary perturbations may significantly affect the planet growth (see for example \cite{marzari2000,nelson2000,thebault2004,thebault2006,quintana2007,haghi2007, paar2008,muller2012,marzari2012,thebault2015}) but, at present, it is not fully understood which may be the final characteristics of  planetary systems in binaries.  However, the presence of a massive planet in an S--type orbit for example around $\gamma$ Cephei ($m_p \sim 1.7 m_J$), HD 41004 A ($m_p \sim 2.3 m_J$) and Gliese 86  ($m_p \sim 4 m_J$), which are close binaries with semi--major axis of about 20 au and eccentric orbits, and the detection of many others, even multiple planetary systems, as illustrated in Fig. 1 of \cite{galax2019}, suggest that the growth of massive planets may be common in binaries.  The formation of a system of three giant planets, even in binaries with small $a_B$, may occur with a frequency close to that around single stars. 

One critical issue when exploring the dynamical evolution of planets in binaries is the variety of possible dynamical configurations due to the large number of free parameters like the semi--major axis, eccentricity and inclination of the binary orbit and the mass ratio of the two stars. Hereinafter, we will adopt a configuration in which the binary is made of two equal mass solar--type stars and we will look for scaling relations that may reduce the number of degrees of freedom. 

\section{Scaling of the dynamics with the ratio \texorpdfstring{$R = a_1 / a_B$}{R}} 

In order to reduce the number of free parameters in the dynamical exploration  of three planets in a binary star system,  we have first looked for possible scaling relations with some representative (canonical) parameters. The most appealing one appears to be the ratio between the semi--major axis of the binary and that of the inner planet $R = a_1 / a_B$.  This scaling is relevant considering that  a binary companion close to the primary star truncates the circum--primary disk reducing the space and mass available to form planets, in particular three giant planets.  However, if the dynamical results depend on $R$, the outcomes on stability found for a given $a_B$ can be scaled for smaller or wider binaries. The reliability of this scaling is related to the fact that resonances between the planets and  with the binary companion depend only on the ratio between $a_B$ and $a_i$ and not on their  absolute values. An additional confirmation can be found in the linear dependence of the critical semi--major axis $a_c$ for a single planet stability with $a_B$ \citep{HW1999}. This is discussed in more detail at the end of next Section. 

\subsection{Scaling of the instability times with \texorpdfstring{$R$}{R} and the keplerian period}

To better outline the meaning of the $R$ scaling, we have performed a series of numerical simulations where we have computed the timescale for the onset of instability for different values of the planet and binary separation.  As usually done in these kind of explorations (see for example \cite{chambers1996, marzari2002,chatterjee2008}), to evaluate the increase in the time interval for orbital crossing as a function of the planetary separation we have scaled  the initial distance between each pair of planets with the mutual Hill radius $R_H$.  The initial semi--major axes of the planets are defined as $a_1$, $a_2 = a_1 + K R_H$ and $a_3= a_2 + K R_H$. In general, for a pair of planets $i$ and $i+1$, the mutual Hill radius is defined as follows
\begin{equation}
\label{eq:rh}
R_H = 
\left(\frac{m_i+m_{i+1}}{3 m_0}\right)^{1/3} \frac {(a_i+a_{i+1})} {2} , 
\quad i=1,2,\ldots,
\end{equation}
where $m_0$ is the mass of the primary, and $m_i$ are masses of the planets.

We consider co--planar orbits for all planets with initial eccentricities randomly selected between 0 and 0.01 and fully random initial orbital angles. We focus on giant planets with $m_p = 1 m_J$ and super Earths with $m_p = 10 m_{\oplus}$.  The simulations are stopped after the first close encounter between two planets, which marks the onset of instability, or at the maximum time of  $t=10^8$ yrs.  The numerical integrator RADAU  \citep{radau1985} has been used to compute the orbital evolution of the planets and the binary companion.

\begin{figure}
\centering
\includegraphics[width=\columnwidth]{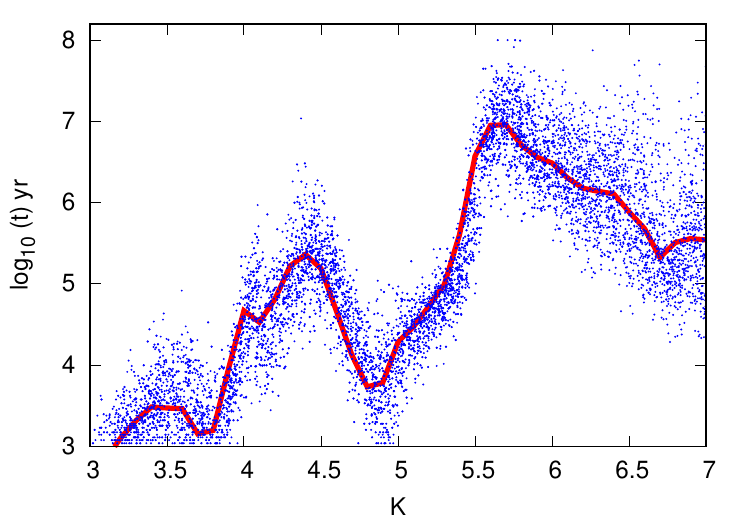}
\caption{Timescale for the instability onset in a configuration with $a_1= 2$ au, $a_b=50$ au ($R=0.04$) and $e_b=0.4$. The blue dots represent the actual numerical data while the red line is the outcome of the averaging over each bin in $K$. }
\label{fig:fit}
\end{figure}

\begin{figure}
\centering
\includegraphics[width=\columnwidth]{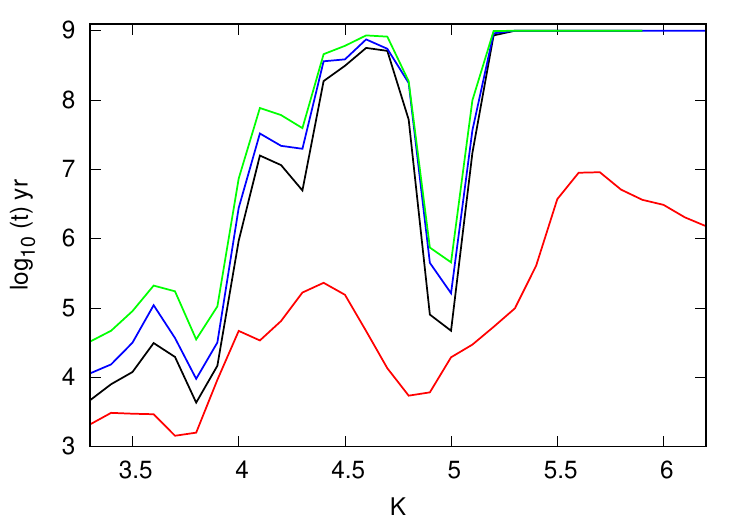}
\caption{{Timescale of the onset of instability for three planets orbiting a single star with  $a_1= 3$au (black line), $a_1= 6$ au (blue line) and $a_1= 12$ au (green line), respectively. For values of $K$ larger than 5.3, all systems are stable over timescales longer than $10^9$ yrs. As a comparison, we have also added the red line marking the instability times for the close binary case shown in Fig.~\ref{fig:fit} } }
\label{fig:3pla_sca}
\end{figure}

\begin{figure}
\centering
\includegraphics[width=\columnwidth]{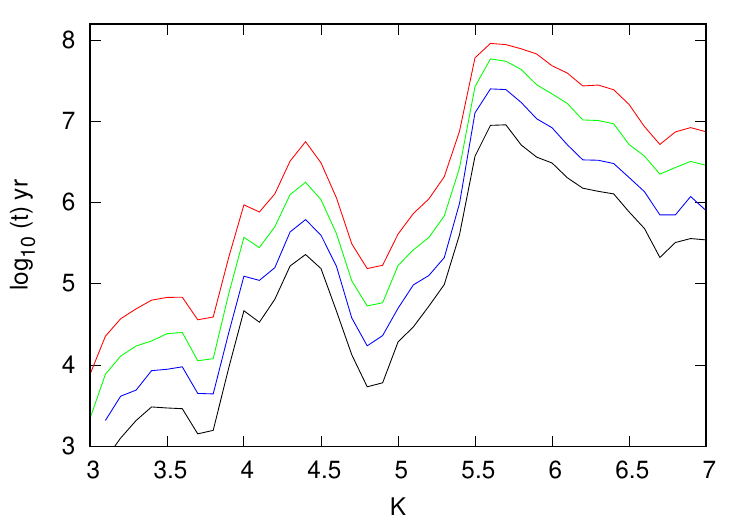}
\includegraphics[width=\columnwidth]{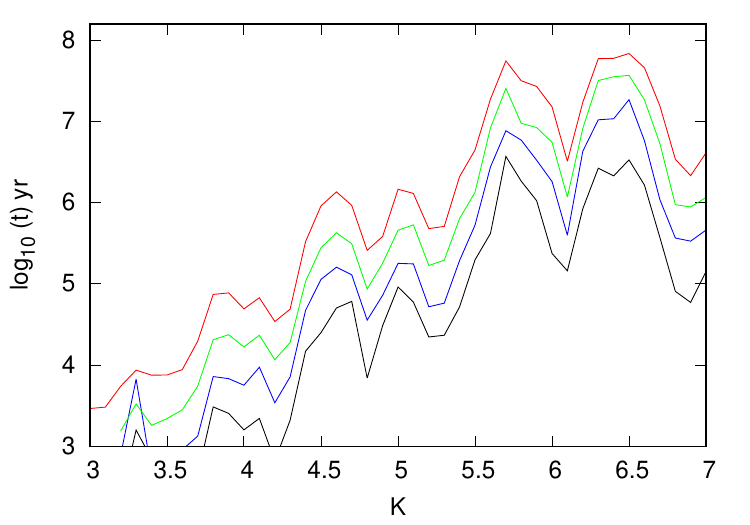}
\caption{Average instability times for Jupiter size planets (top panel) and super--Earths (bottom panel). The  continuous black line is for $a_B = 50$ au, the blue one for $a_B = 100$ au, the green one for $a_B = 200$ au and the red one for $a_B = 400$.  In all models $R=0.04$ and $e_B = 0.4$. }
\label{fig:aB_scale}
\end{figure}

We consider different initial configurations where we fix $a_1$ and $a_B$ (and therefore their ratio $R$), and we perform a large number of simulations with random initial values of $K$. To better compare the outcome of the numerical results for different values of $R$, all the instability times are grouped in bins in $K$ and on each bin the average instability time is computed. An example is shown in Fig.\ref{fig:fit} where the actual data on the timescale for the instability onset vs. $K$ are shown together with the averaged logarithmic fit represented by a continuous line.   
On purpose, we select a dynamical configuration which is chaotic for any value of $K$ in order to see the changes in the instability time for different values  of $R$. If the system is analysed on an interval of $K$ values for which it is stable, then we will have all the times saturated to $10^8$ preventing the identification of a trend with $R$. It is also interesting to note that the same initial configuration for three planets around a single star has stability times longer than $10^9$ yr for $K \geq 5.3$ {as shown in Fig.~\ref{fig:3pla_sca} and illustrated also in \citep{marzari2002}. The sharp drop in the stability times around $K= 3.8-4$ is due to the presence of the 5:3 resonance between the first and second planet and the 3:1 resonance between the first and third planet while that at $K = 5$ is due to the 2:1 resonance (see \cite{marzari2002,Raymond_2010}.   } 
In our binary configuration, {long stability times are never achieved because} when $K$ becomes large enough to grant the stability of the planets against their mutual perturbations, the binary perturbations come into play preventing the onset of long--term stability (Fig.\ref{fig:aB_scale}). 

In Fig.\ref{fig:aB_scale}, top panel, we compare the averaged time--scales computed for different configurations with $a_B = 50,100,200,400$ au and $a_1=2,4,8,16$ au, respectively. All these models share the same value of the scaling parameter $R = 0.04$. Note that the average timescale may be slightly underestimated close to $10^8$ yrs due to the saturation of the instability times to $10^8$ yrs, our limiting integration time--span.

The trend in Fig.\ref{fig:aB_scale} is clear: the curves follow the same pattern as a function of $K$ in the 4 different models and the value in each bin is scaled by the keplerian period. Each curve can be superimposed almost perfectly to the next by scaling the values with $2^{3/2}$ (2 is the ratio between the subsequent values of $a_B$ and $a_1$), starting from the case with smaller $a_B$ and $a_1$. {This behaviour is observed also in Fig.~\ref{fig:3pla_sca} in the case of three planets around a single star.}

The scaling trend with the Keplerian period is observed in both the models with Jupiter--size planets ($m_p= 1 m_J$) in the top panel
of Fig.\ref{fig:aB_scale}
and in the case of super--Earths ($m_p=10 m_{\oplus}$) in the bottom panel.  This behaviour  strongly suggests that the dynamics, once fixed the ratio $R$, is the same for different initial values of $a_1$ and $a_B$.

The scaling property may be justified based on the form of the $N$-body equations of motion. In the barycenter frame they read as follows
\begin{equation}
\label{eq:nbody}
\ddot{\vec{r}}_i = k^2 \sum_j m_j \frac{\vec{r}_{ji}}{|{\vec{r}}_{ji}|^3}, 
\quad i\neq j=0,\ldots,N,
\end{equation}
where ${\vec{r}_{ji}} \equiv {\vec{r}}_j - {\vec{r}}_i$ is the relative radius vector from a body $i$ to a body $j$, $j=0$ denotes the primary star $m_0 \equiv m_{\star}$, $m_i,m_j$ are for the planet masses and the secondary, and $k^2$ is the gravitational constant. The scaling invariance of the ODE system~(\ref{eq:nbody}) means that if a particular $\vec{r}_{i}(t)$ is the solution to these equations, then $\rho^{-2/3} \vec{r}_i(\rho t)$ with some scaling factor $\rho>0$ is also the solution.  As the result, the orbital radii, eccentricities and relative phase angles in  geometrically re-scaled copies of the system exhibit the same dynamical character as the original system. Since linearly rescaled orbits evolve such as $\sim \vec{r}_i(\rho t)$, the time-scale of the orbital evolution changes as $t \rightarrow \rho t$. A particular example and an illustration of the $N$-body scaling may be found in \cite{Gozdziewski2018}, as derived for a system of four massive Jovian planets, see their Fig.~4.

In order to explain the results in Fig.~\ref{fig:aB_scale}, we note that the semi-axes distribution $a_{i+1} = a_i + K R_{H}$ with $R_H \equiv \beta(m_i,m_{i+1},m_0) (a_i+a_{i+1})$, as given in Eq.~\ref{eq:rh}, leads to
\[
a_{i+1} = a_1 \gamma^{i+1}, \quad
\gamma \equiv \frac{1+K \beta}{1-K \beta}.
\]
Since $\gamma$ is constant for fixed $K$, and equal planet masses $m_i$, the planetary system scales linearly with $a_1$. Now, if to take $\rho^{-2/3} \simeq 2$, as in Fig~\ref{fig:aB_scale} we considered the semi-major axes doubled between subsequent system copies, then $\rho \simeq 2^{3/2}$, i.e., the factor found above that simultaneously scales the instability (event) time.

\subsection{Invariance of MEGNO stability maps  for the same value of \texorpdfstring{$R$}{R}.}

As an additional test to confirm the reliability of the dynamical scaling with $R$ we have used the fast indicator MEGNO \citep{cincotta2000,goz2001,cincotta2003} to compare the dynamical maps in  two different configurations, in terms of binary separation, but with the same $R=0.005$. The first has $a_B=200$ au and $a_1=1$ au while the second has $a_B=50$ au and $a_1=0.25$ au. In both configurations the binary eccentricity is set to $e_B=0.4$. The exploration of the phase space is performed at random with  $a_1$ fixed by $R$ while $a_2$ and $a_3$ are randomly chosen within the critical semi--major axis $a_c$ computed according to the formula given in \cite{holman1997}. The initial eccentricities are random and lower than 0.01 as before. In both cases the orbits are integrated over 10000 binary orbits. 

\begin{figure}
\centering
\includegraphics[angle=-90,width=\columnwidth]{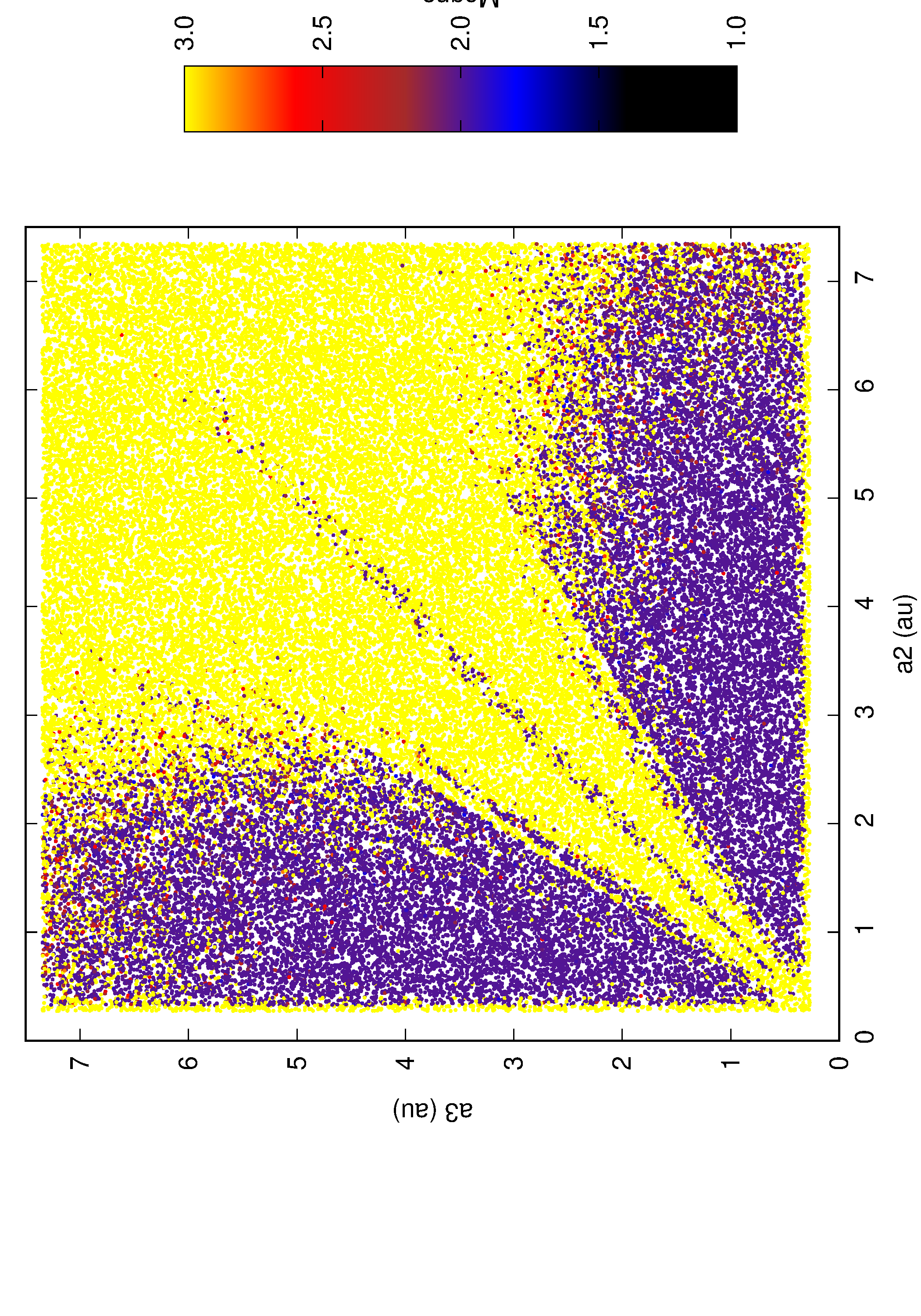}
\hspace{-1cm}
\includegraphics[angle=-90,width=\columnwidth]{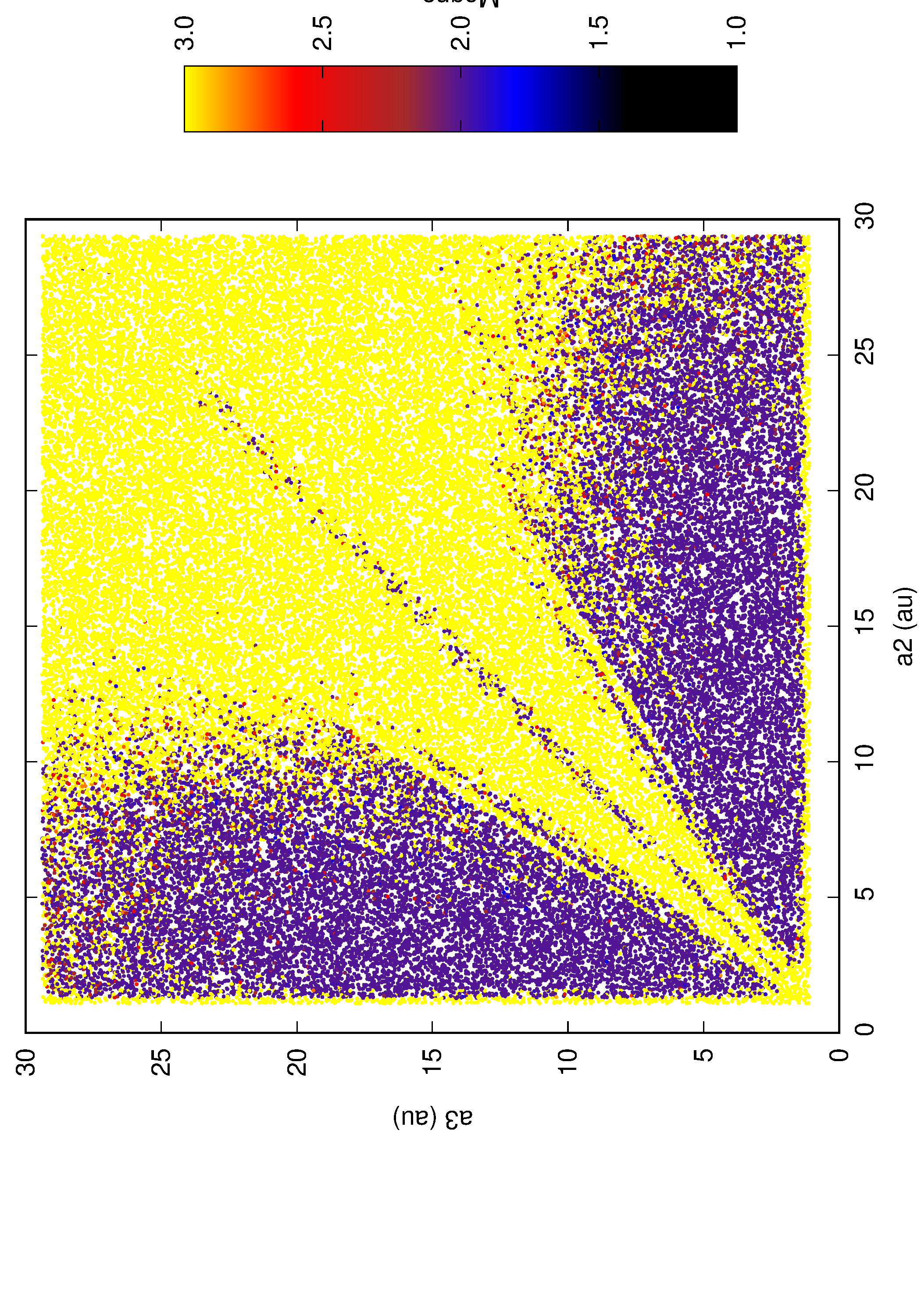}
\caption{Stability maps for the  configuration with $a_B=50, a_1=0.25$ au (top panel) and $a_B=200, a_1=1$ au (bottom panel). The two plots do not show significant differences. The color coding gives the value of MEGNO which, for long term stability, must be $\sim 2$. }
\label{fig:aB_scale_megno}
\end{figure}

In Fig.\ref{fig:aB_scale_megno} we compare the two stability maps where the value of MEGNO is plotted as a function of $a_2$ and $a_3$ ($a_1$ is fixed).  The stable regions, marked by the violet dots, appear indistinguishable one from the other (note the different scale in the semi--major axes). This finding reinforce the validity of the scaling with $R$ even  in terms of formally chaotic/regular evolution in terms of the Lyapunov Characteristic Number (LCN), for time scales characteristic for at least 2-body MMRs. Similarly given the $N$-body ODEs, this property should be preserved in linearly scaled systems also for the time--scales related to the three-body and secular resonances.

\section{Dependence of the planetary system stability on the value of \texorpdfstring{$R$}{R}}

Once found that the dynamics scales with the ratio $R$, it is important to test how the stability of the system, intended as width of the stable regions, depends on the initial choice of $R$. By increasing $R$, we shift the system of the three planets outwards, getting closer to the outer stability limit in semi--major axis $a_c$. It could be guessed that the maximum value of $R$ for which stable regions exist corresponds to the configuration in which we pack all three planets as close as possible in order to meet the Hill's stability criterion and the outer planet has a semi--major axis near to $a_c$. For example, for $a_b = 200$ and $e_B=0.4$, assuming that the minimum value of the factor $K$ for the separation of the three planets for stability is 6 (according to \cite{marzari2002} for $K$ larger than 5.3 the stability timescale is of the order of $10^9$ yr), we would expect a maximum value of $a_1=10.4$ au for packed stable systems which corresponds to $R_0=0.052$.

However, the binary perturbations affect the stability of the three planets in two different ways. First they force an increase in the value of $K$ for stability because the binary companion forces secular perturbations on the planets  \citep{heppenheimer1978,andrade2016,pilat2016} causing  an increase in the eccentricity and leading to a higher number of secular resonances which potentially overlap with those among the planets. In addition, there may be a shrinking in $a_c$ due to the mutual perturbations of the planets.  To test the maximum value of $R$ for stability,  we have analysed systems with the same value of $a_B=200$ and $e_B=0.4$ but with different values of $a_1$ (i.e. $R$),  from 2 to 6 au.  

\begin{figure*}
\centering

\begin{tabular}{cc}
\hspace{-1cm}
\includegraphics[angle=-90,width=\columnwidth]{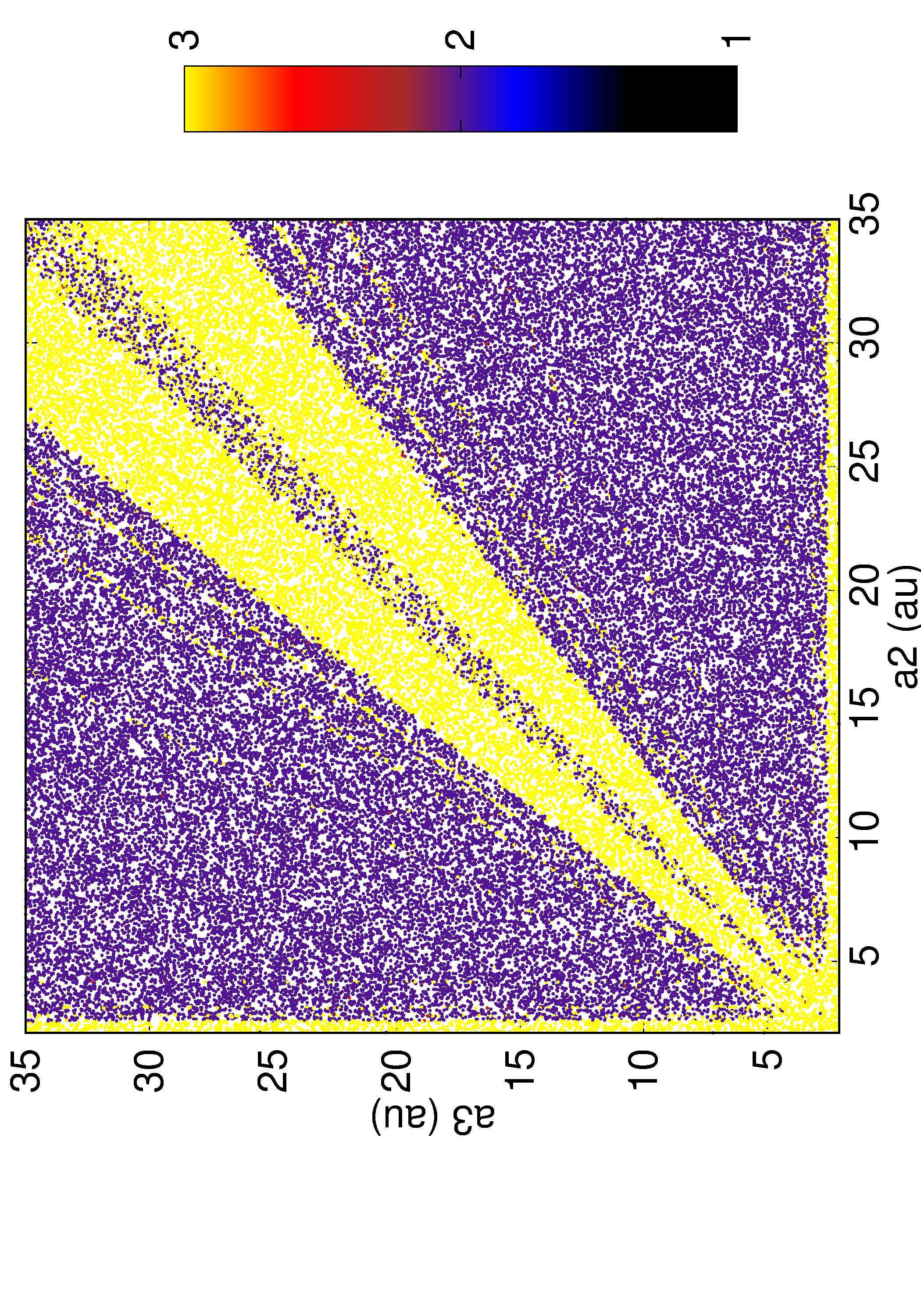} & \hspace{-0.5cm}  
\includegraphics[angle=-90,width=\columnwidth]{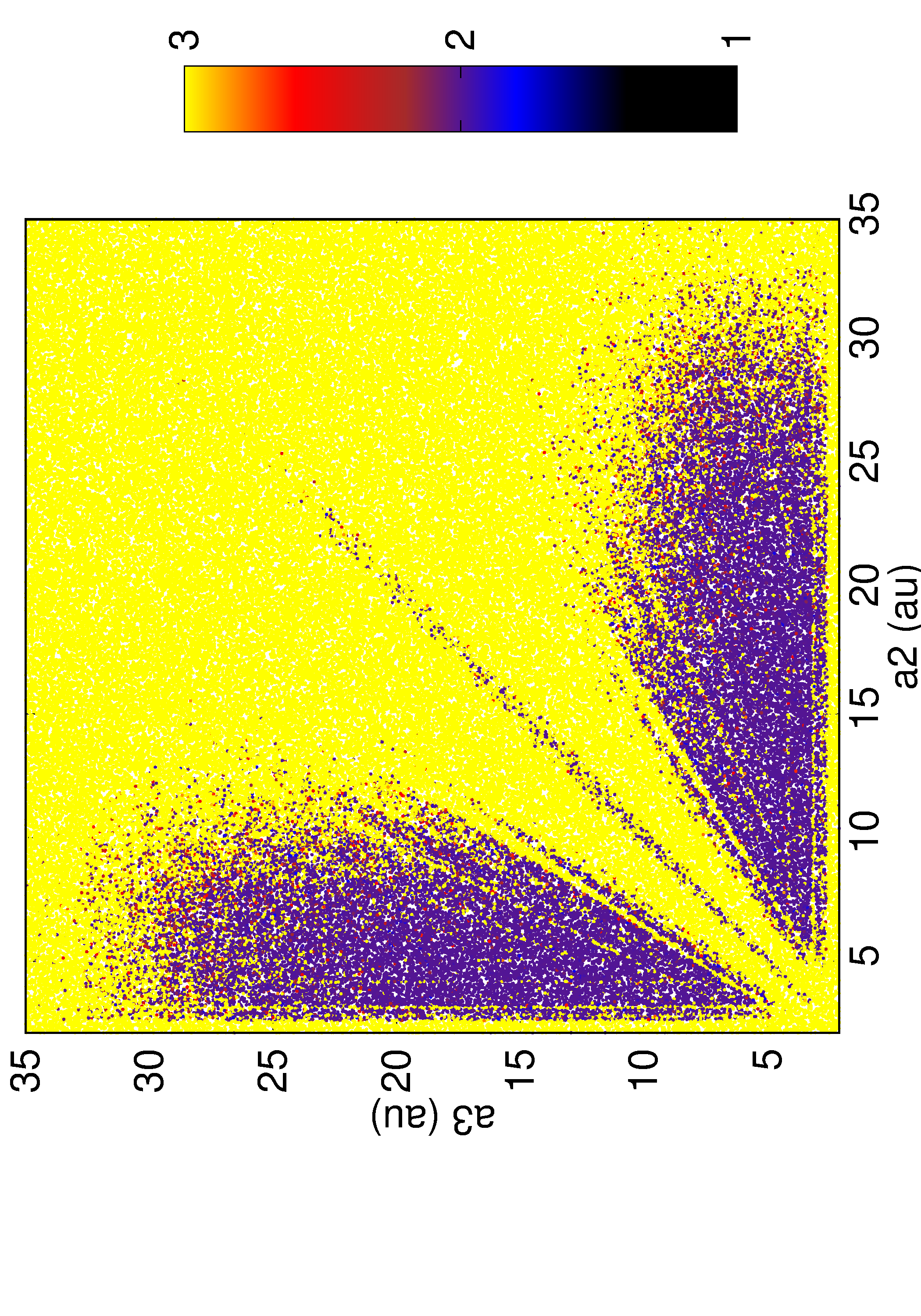} \\  
(a) & (b) \\[6pt] 
\hspace{-1cm}
\includegraphics[angle=-90,width=\columnwidth]{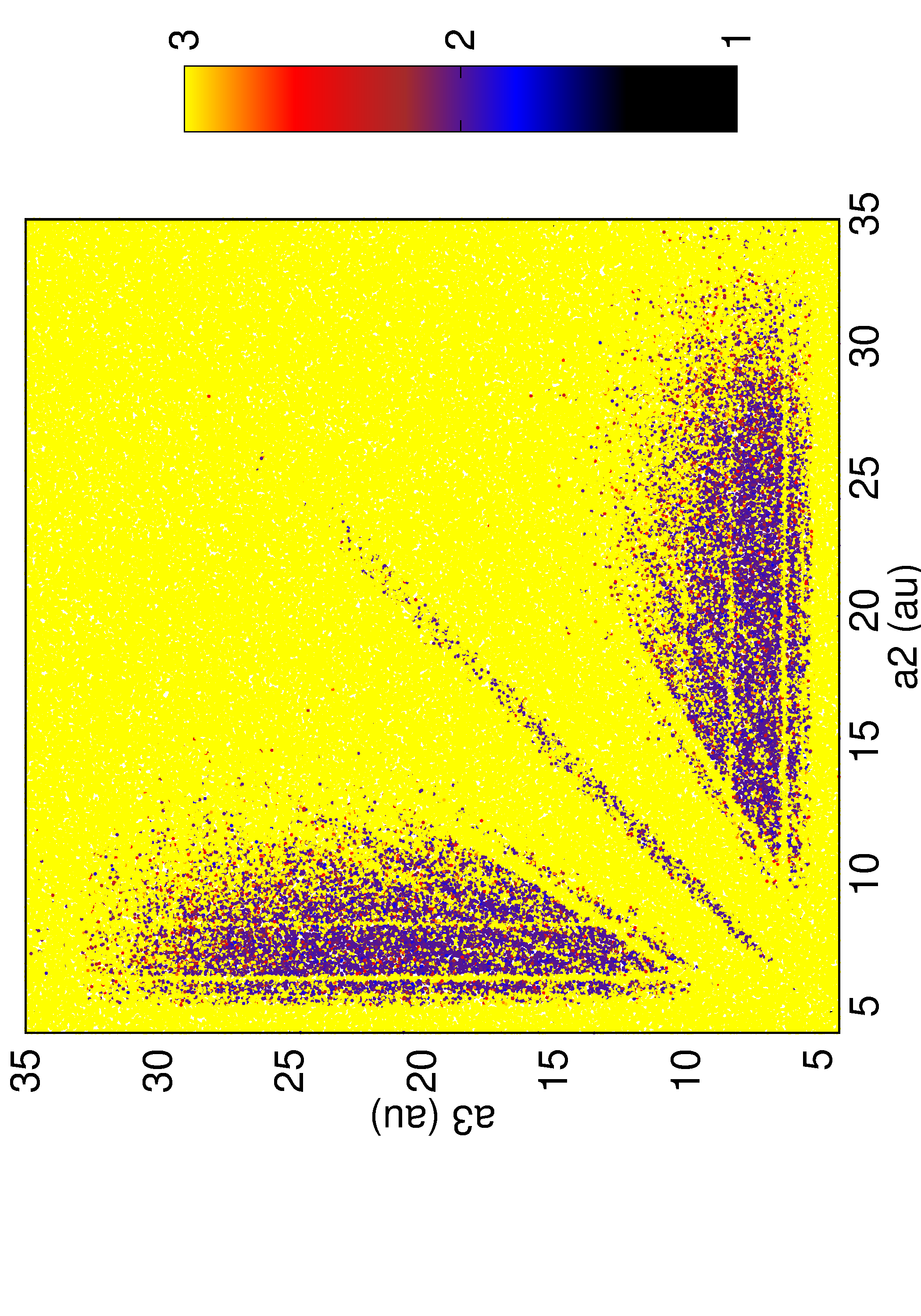} & \hspace{-0.5cm}
\includegraphics[angle=-90,width=\columnwidth]{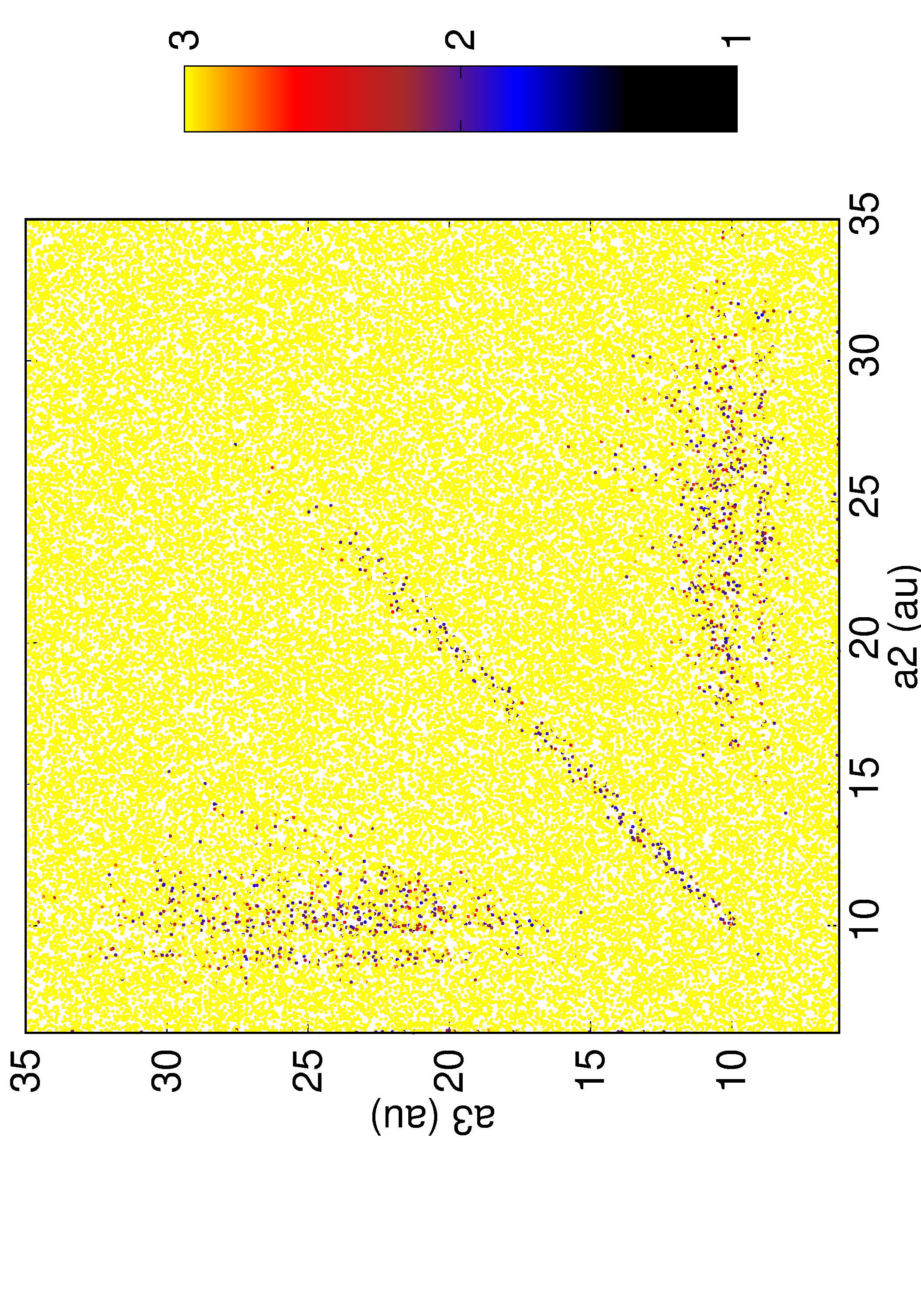} \\
(c) & (d)  \\[6pt]
\end{tabular}
\caption{Stability maps for the case without the binary companion (panel a) and for three different values of $a_1$ for the same binary configuration ($a_B=200 au, e_B=0.4$). Panel b) is for $a_1=2, R=0.01$, panel c) for $a_1=4, R=0.02$ and panel d) is for $a_1=6, R=0.03$. Beyond this last value of $R$ the stability regions have a negligible size.}
\label{fig:stab_maps}
\end{figure*}

In Fig.~\ref{fig:stab_maps}  we show the  stability maps for the case of a single star (top left panel)  and three binary cases with $R= 0.01, 0.02, 0.03$. The stable region in the phase space progressively shrinks due to the binary perturbations and, for $R \geq 0.033$, there are no stable configurations for three planets. This is significantly lower than the values of $R_0 = 0.052$ which was previously derived by simply packing planets within $a_c$ and neglecting the perturbations of the binary. From a physical point of view, this implies that there cannot be systems of three planets, for $e_B = 0.4$, where the inner one is farther from the star than $a_1=R \cdot a_B$ where $R=0.033$. This does not imply that they could not have formed farther out because, for example, they might have grown at different times and then migrated inwards by interaction with the disk. 
 
 \section{Dependence of the stability limit on the binary eccentricity \texorpdfstring{\lowercase{$e_B$}}{eB}}
 
\begin{figure}
\centering
\includegraphics[angle=-90,width=\columnwidth]{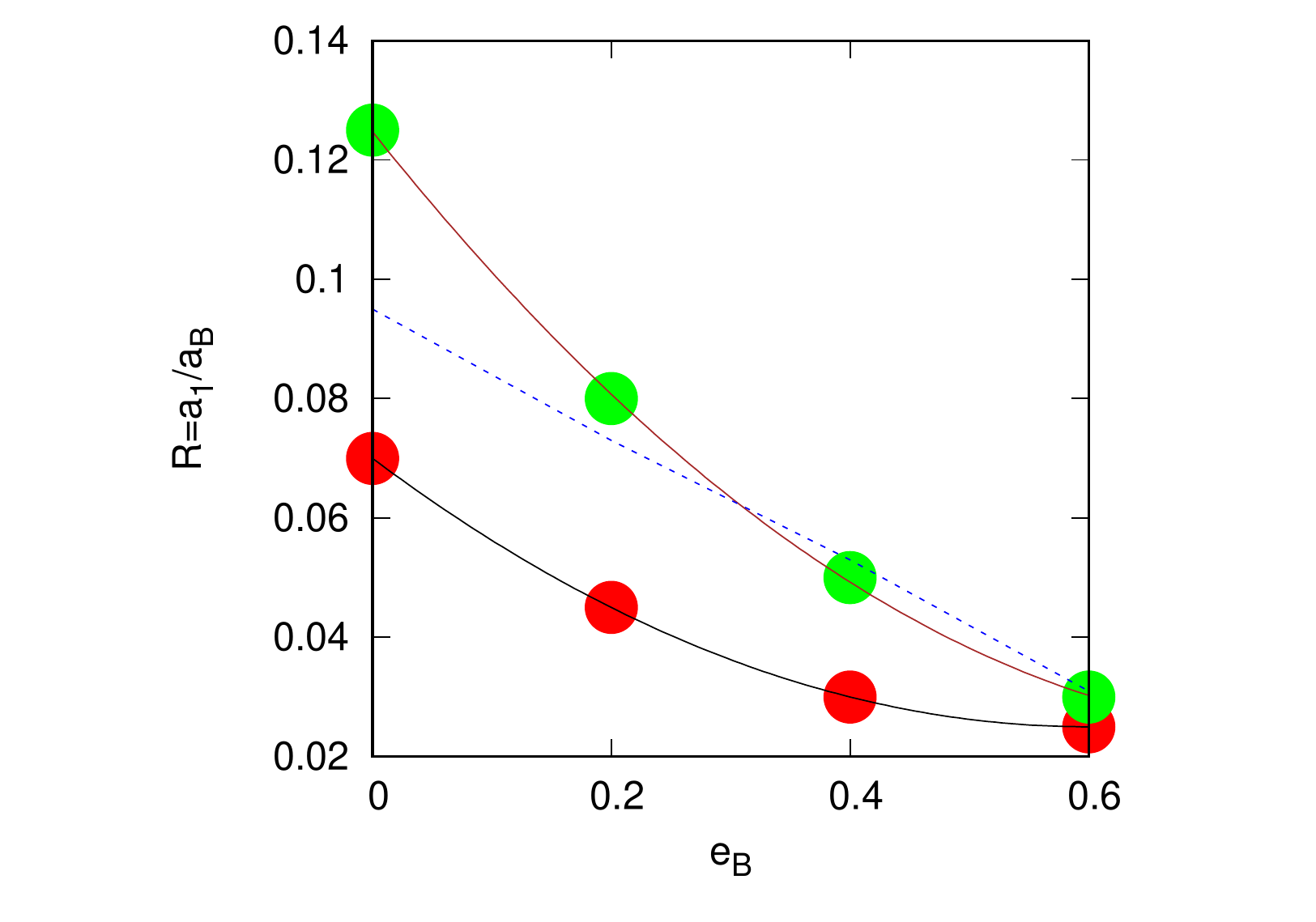}
\caption{Dependence of the ratio $R$ over the binary eccentricity $e_B$.  The red filled circles (and relative fitting line) refer to Jupiter--size planets while the green filled circles refer to $m_P = 10 M_{\oplus}$ Neptune--size planets. The blue dashed line show the prediction for the stability of a three planet system ($m_p = 1 M_J$) packed within $a_c$ and Hill's stable if around a single star. }
\label{fig:fit_ecce}
\end{figure}

 To test how the the maximum value of $R = 0.033$ for stability depends on the binary eccentricity $e_B$, we have performed a new set of simulations where we compute MEGNO for different values of $e_B$. When the stability map is similar to that shown in Fig.\ref{fig:stab_maps}, panel (d), i.e., when less than 0.1\% of the sample systems are stable with MEGNO$\simeq 2$, we assume the corresponding value of $R$ as threshold value beyond which the system of three planets is always unstable. The outcomes of the simulations for $e_B = 0.0,0.2,0.4,0.6$ are shown for $m_p = 1 M_J$ (red filled circles) and $m_P = 10 M_{\oplus}$ (green filled circles) in Fig.~\ref{fig:fit_ecce}. There is a well defined trend with $R$ decreasing for higher values of $e_B$. Performing a  polynomial fit to the data, we get the following semi--empirical formulas:
 
 \begin{equation}
R_J = (0.070 \pm 0.001) - (0.150 \pm 0.002) e_B + (0.125 \pm 0.004) e_B^2
 \end{equation}
 
for Jupiter size planets, and

\begin{equation}
R_N = (0.125 \pm 0.001) - (0.251 \pm 0.008) e_B + (0.156 \pm 0.013) e_B^2    
\end{equation}

for Neptune size planets. 

The fits are very precise and illustrate the growing instability limit with $e_B$. For $e_B=0.4$, we retrieve from the first fit the value of $R$ computed in the previous section for three Jupiter--size planets.  The blue dashed line illustrates the analytical prediction obtained by simply packing the planets within the critical $a_c$ \citep{holman1997}  and according to Hill's stability criterion. 
The significant difference between the two curves strongly suggests that the mutual gravitational perturbations among the planets contribute to shrink the stability region that would be available in presence of the binary perturbations only. Secular perturbations, secular and mean motion resonances among the planets affect the stability, even superimposing to that due to the secondary star, creating additional chaotic regions and reducing the stable areas. 
The same trend is also observed  for less massive planets and it confirms that the stability of the planetary system is very sensitive to the binary eccentricity.

We can conclude that the stability properties of a systems of three planets in a binary scale with the parameter $R$ and the threshold stability value strongly depends on the binary eccentricity. These results concern the stability of the planetary systems, the question we have to answer now is how reliable is the $R$ scaling when we consider the outcome of P--P scattering events which include additional physics respect to the pure $N$--body case analysed so far. 

\section{The numerical model for the P-P scattering}

To investigate the outcome of P--P scattering  we use a fourth-order time-symmetric Hermite code \citep{Kokubo98} with a variable shared time step. In the numerical algorithm we include the effects of tides on the planet, mutual collisions, and GR \citep{marzari_nagasawa2019}. The tidal interaction between a planet and the central star is a complex problem since it includes both static tides, where frictional processes within the planet remove energy from its orbit, and dynamical tides where the energy dissipation occurs through the excitation and damping of oscillations \citep{ivanov2004, ivanov2007}. In our model we include dynamical tides \citep{naga08} related to the fundamental modes and inertial modes of a planet in a state of pseudo-synchronization as described in \citep{ivanov2007}.  We neglect the contribution from dynamical tides raised by  the planet  on  the  host  star since, according  to
\cite{marzanaga2019}, they are weaker and the planetary fundamental modes dominate the evolution of the system. 
In our simulations we compute the effects of the dynamical tides with the impulse approximation given by \cite{ivanov2007}.  However, this approach  becomes less accurate when the  circularization  proceeds  (e.g.,  \cite{mardling1995a,mardling1995b})   and  the eccentricity becomes low. In this regime, equilibrium tides become effective (e.g., \cite{beauge2012}) and the evolution occurs on longer timescales. It is a complex task to predict when dynamical tides give way to the equilibrium ones and we prefer to continue the numerical integration with dynamical tides even for low eccentricities but we stop our simulations when  the  energy,  decreasing  from  the  tide  at  the  pericenter overcomes the orbital energy. This leads to a clustering of tidally circularized planets around 0.02 au \citep{naga08, Nagasawa2011, marzanaga2019}.

When two planets get closer than the sum of their radii, we assume that a collision occurs and the two bodies merge into a single body with twice the mass. The new velocity vector of this single body is derived from the conservation of momentum while its radius is computed assuming the same density of the colliding planets.  When a planet hits one of the star it is removed from the simulation while those that are fully circularized are no longer evolved in time. 

In these simulations we abandon the coplanar configuration and we assume that the planets have a small random initial inclination lower than $3^o$ respect to the binary orbital plane. 

\section{P--P scattering outcomes}

An unstable 3--planet system may end up in different final configurations. These include: 
\begin{itemize}
    \item Ejection of one  planet from the system (this includes impact on either stars). One planet is left on a close orbit and the two surviving bodies are on a dynamically stable configuration 
    \item Two planets are ejected and a single planet survive on a close orbit
    \item Two planets collide and a stable two--planet configuration is obtained where one planet has twice the mass of the other
    \item In all previous cases the inner planet can be tidally circularized on a short timescale to a orbit very close to the central star. 
    \item  In all previous cases an inner planet in a Kozai state with the companion star is produced.
\end{itemize}

\begin{figure}
\centering
\includegraphics[angle=-90,width=\columnwidth]{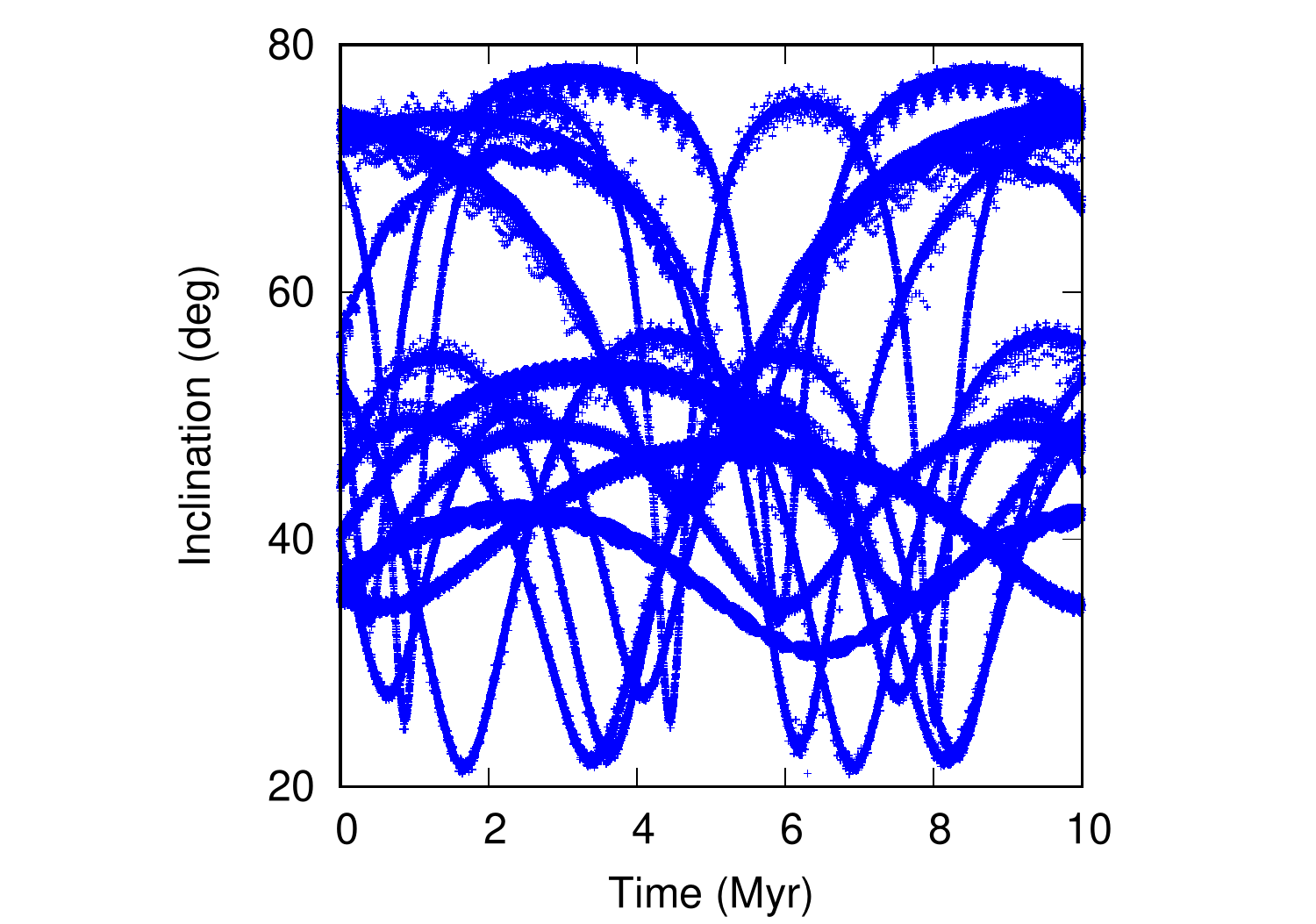}
\includegraphics[angle=-90,width=\columnwidth]{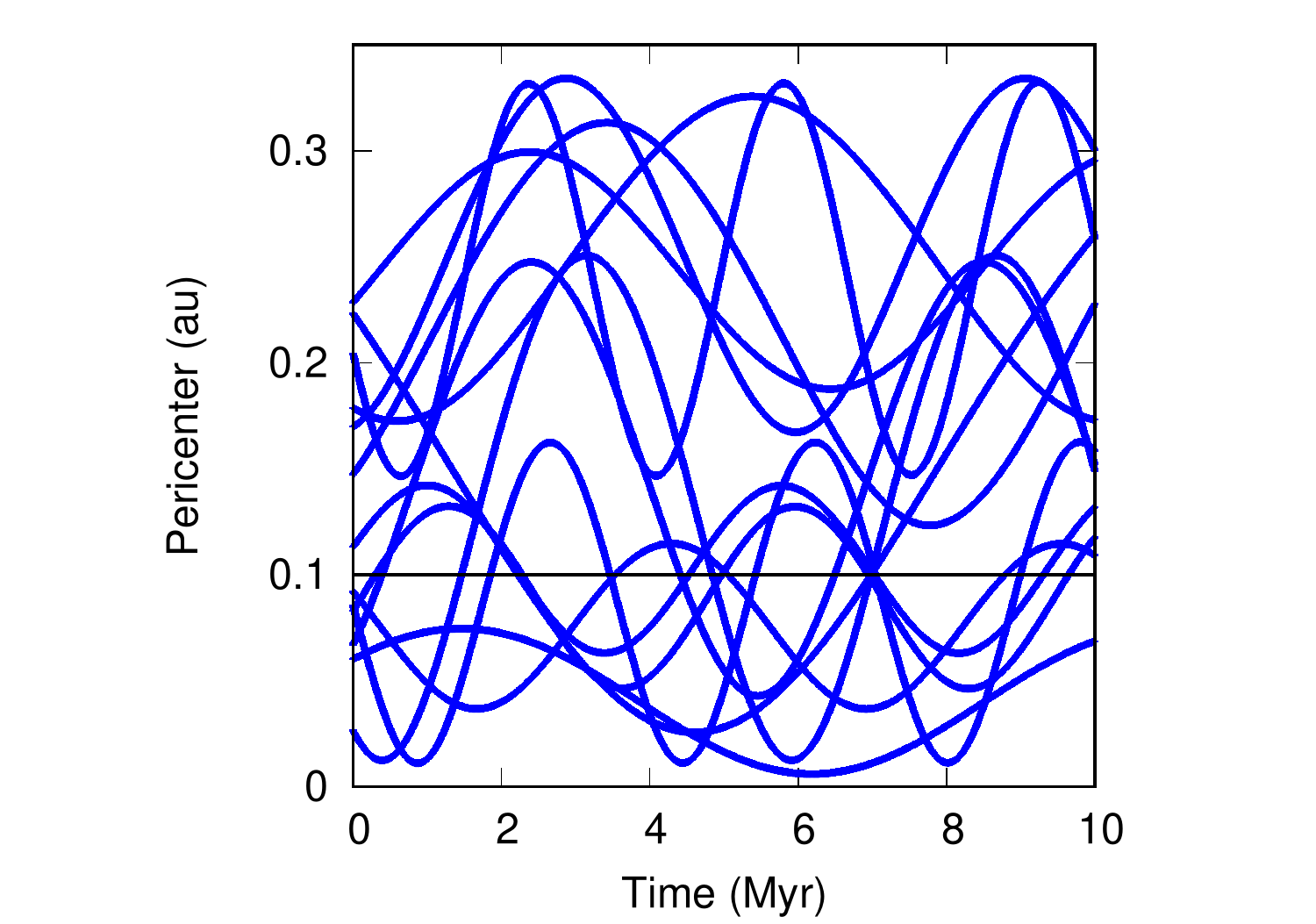}
\caption{Evolution of the inclination (top panel) and pericenter distance (bottom panel) in cases which are in a Kozai state when $e_B=0.4$ and $a_B=200$au.}
\label{fig:kozai}
\end{figure}

The formation of a Kozai state is tested with an additional $N$-body simulation for 10 Myr (which is longer than the estimated Kozai period in out binary configuration according to \cite{naoz2016}). In 
Fig.\ref{fig:kozai} the evolution of some of these systems is shown vs. time for $e_b = 0.4$, $a_b =200$ au. In a fraction of cases the pericenter remains larger than 0.05 au all the time in spite of the high eccentricity achieved during a Kozai cycle and these may be configurations similar to that of 16 Cyg b \citep{holman1997}. Those achieving a pericenter smaller than 0.05 au can be tidally circularized into Hot/Warm Jupiters on a long timescale due to slow or fast migration \citep{Petrovich2015} and end up with a semi--major axis smaller than 0.1 au. These are the most interesting cases that will be recorded in the analysis.

\section{Final orbital distribution in a test case}

\begin{figure}
\centering
\includegraphics[width=\columnwidth]{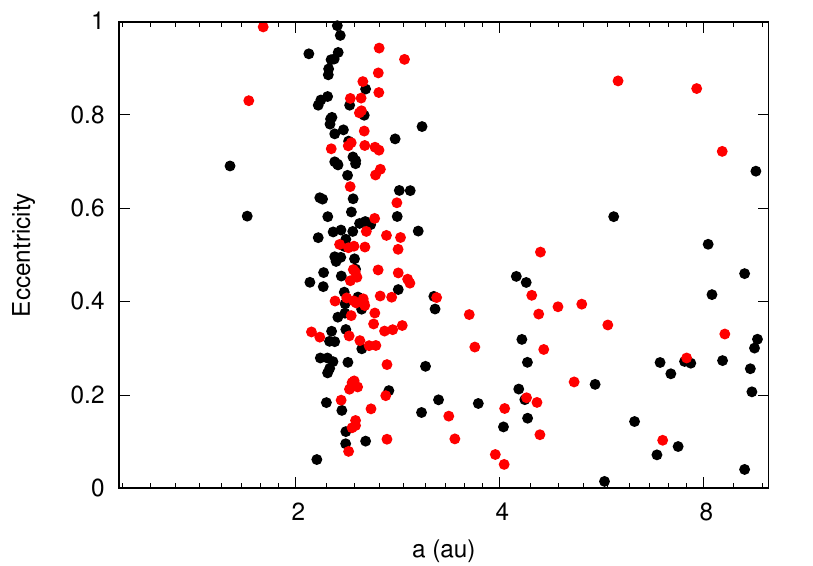}
\includegraphics[width=\columnwidth]{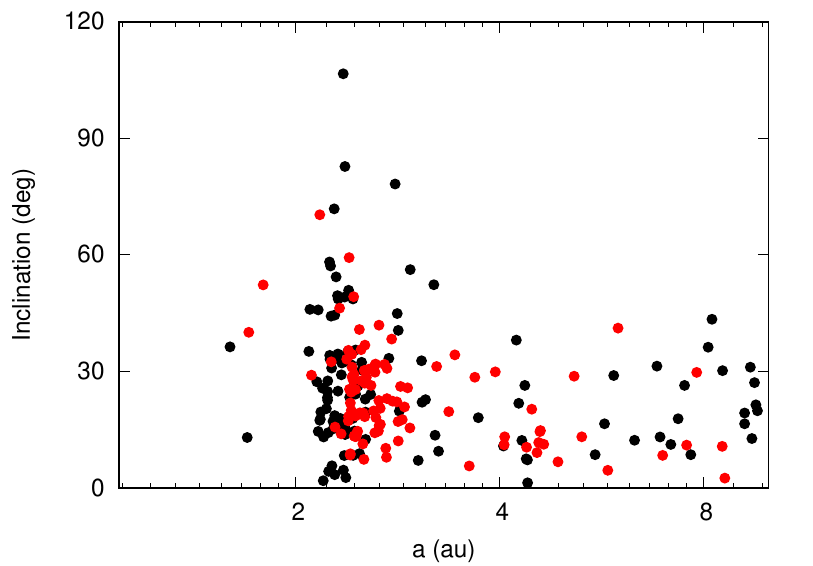}
\caption{Final distribution of the orbital elements after 2 Myr of integration. The black filled circles mark the model without the binary while the red ones are for the binary case with $a_B=200$ au and $e_B=0.4$.}
\label{fig:distri}
\end{figure}

We first perform a comparison between the end state of P--P scattering with and without the binary companion. A configuration where the inner planet has
$a_1 = 5$ au is considered in both cases while for the binary case the companion orbit has $a_B=200$ au and $e_B=0.4$.  The semi--major axes of the second and third planet are computed as  $K R_{\corr{H}}$ where $K$ is set to 3.5. This value gives a short timescale for the onset of instability, of the order of $10^4$ yrs, allowing a good statistical sampling of the P--P scattering events on a reasonable CPU load.  At the end of the two numerical integrations, lasting  10 Myr, we compare the final orbital elements of the surviving planets. In this first test, we neglect those which are tidally circularized and we focus on the final orbital distribution to check the potential influence of the companion star in the distribution of the eccentric planets. According to Fig.\ref{fig:distri}, there is a higher dispersion in semi--major axis in the binary case (red filled circles) compared to the case around a single star (black filled circles). 
In the latter, the inner planets are well aligned to the semi--major axis value which is predicted by the conservation of the orbital energy for three planets. In the presence of the binary (red filled circles), the semi--major axis of the inner planets is slightly shifted outwards and more dispersed.
This is expected because the stellar companion absorbs part of the energy of the escaping planets which, as a consequence, is not fully released to the surviving planets, in particular the inner one. This explains why the two distributions are different. In terms of eccentricity and inclination distribution, the P--P scattering between the planets dominate and there are not significant differences in the two cases. 

For smaller values of $a_1$, i.e. 1 and 3 au, the difference between the case without binary and that with the inclusion of the binary companion is less marked because there are more encounters between the planets which may interact with the companion star only after a number of encounters between them that is higher compared to the case with $a_1 = 5$ au. However, the outputs of the chaotic evolution are, even for small initial values of $a_1$, statistically different, as shown in the next Section.  

\section{Statistical results}

We present in this section statistical estimates of the occurrence of distinct P--P scattering events for different values of the binary orbital parameters. 

\subsection{Fraction of systems with two surviving planets}

To explore the effects of the binary companion on the outcome of the P--P scattering we have run different models with and without the secondary star and adopting different values of the semi--major axis of the inner planet (i.e. different values of $R$). The semi--major axes of the second and third planet are computed as before with $K$ set to 3.5 and the planets are Jovian--size planets.  For each configuration, we run 150 different cases with random initial angles and small initial random values of eccentricity ($e < 0.01$) and inclination ($i < 3^o$). Depending on the semi--major axis of the inner planet, we select a different time--span for the integration: for $a_1 = 1$ au the integration time is 2 Myr, for $a_1 = 3$ au 5 Myr and for $a_1 = 5$ au 10 Myr. In all cases, at the end of the runs the surviving planets are on stable orbits.

\begin{figure}
\centering
\includegraphics[width=\columnwidth]{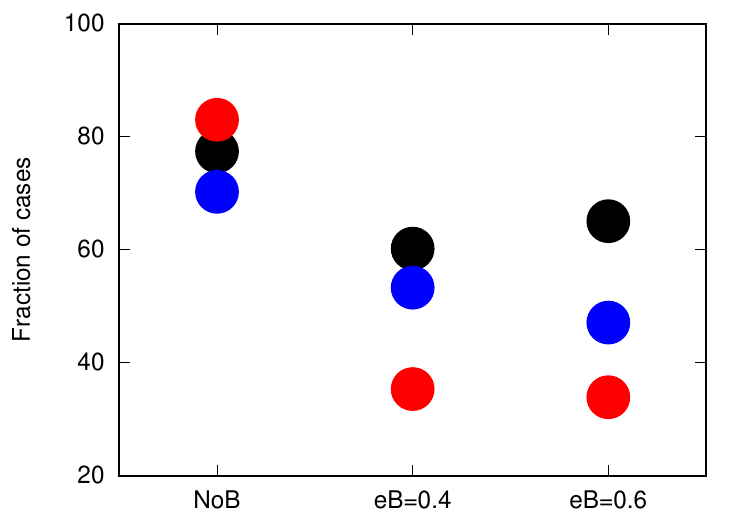}
\caption{Fraction of cases leading, after the chaotic phase, to a two--planet system. The black filled circles are for $a_1 = 1$au, the blue ones are for $a_1 = 3$au and the red ones are for $a_1 = 5$au.  }
\label{fig:2pl}
\end{figure}

In Fig.\ref{fig:2pl} the fraction of cases leading to a final system made of two planets is computed for different values of  $a_1$ and binary eccentricity $e_B$. In the remaining cases, either a collision or the ejection out of the system leave a single planet in the system.  In the figure, we compare the cases without the secondary star (NoB) with two different binary configurations, one with $e_B=0.4$ and one with $e_B = 0.6$. In those without a binary companion the fraction of two--planet systems surviving the chaotic phase ranges from 70\% to 85\% and there is little difference between the three initial values of $a_1$. 

A significant decrease in the number of two--planet systems occurs for eccentric binary systems, as expected according to \cite{marza2005ApJ}. It is noteworthy that in \cite{marza2005ApJ} the binary parameters were different since they adopted $a_B = 50$ au and a mass ratio $\mu=0.4$. This explains the differences observed between their results and the outcome shown in Fig.\ref{fig:2pl} where there is not a significant difference between the case with $e_B=0.4$ and $e_B=0.6$. An interesting feature in Fig.\ref{fig:2pl} is the dependence of the fraction of 2--planet systems on $a_1$. When $a_1=1$ au (black filled circles) the fraction decreases from about 80\% to 60\%. For larger values of $a_1$ the drop is more marked for both  the cases with $e_B=0.4$ and $e_B=0.6$  where the fraction goes down to 50\% when  $a_1 = 3$ au (blue filled circles) and about 30\% when $a_1=5$ au (red filled circles). 

The dependence on $a_1$ of the fraction of planets which are ejected from the system is related to the slower evolution of the cases starting with a larger value of $a_1$. The P--P scattering events occur on a timescale which is expected to be proportional to the Keplerian period. As a consequence, the chaotic  evolution dominated by mutual close encounters between the planets  is faster for a smaller value of $a_1$. During the dynamically violent phase, the planets may achieve high eccentricities and have encounters with the secondary star. The longer the chaotic period, the higher is the probability for a planet to be ejected out of the system by the companion star. As a consequence, the initial location of the planets at the moment in which the instability switches on is comparable, in terms of influence on the final outcome of the P--P scattering, to the binary eccentricity. This may have important implications in the evolution of planetary systems in binaries. If planets form far away from the primary star and undergo little migration, then a predominance of stable single planet systems is expected to be found in binaries.

\subsection{Fraction of planetary collisions}

\begin{figure}
\centering
\includegraphics[width=\columnwidth]{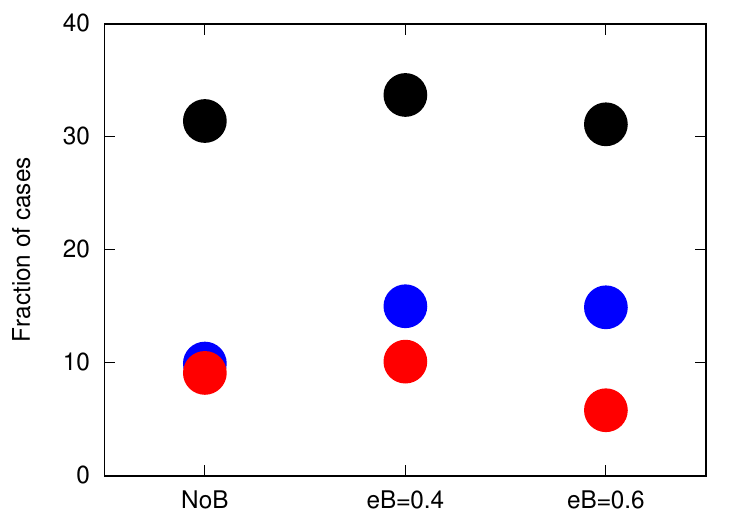}
\caption{Fraction of cases in which two planets collide and merge in a single body. The color coding is the same as in  Fig.\ref{fig:2pl}}.
\label{fig:coll}
\end{figure}

During the chaotic evolution, two planets can collide and merge into a single more massive planet. According to Fig.\ref{fig:coll} the frequency of collisions is not significantly affected by the presence of a companion star and the percentage of collisions appear to be approximately the same even for a highly eccentric binary i.e. $e_B=0.6$. Even in this case, the relevant parameter is the initial distance from the star of the planets, in other words it depends on $a_1$. Smaller initial values of $a_1$  leads 1) to a reduction of the Keplerian period and a higher encounter rate and 2) to an increase of the impact cross section respect to the available space for the orbital evolution.  The combination of the two effects lead to a higher fraction of collisions for small values of  $a_1$ but this does not seem to strongly depend on the presence of the binary companion. In the remaining fraction of events, scattering with ejection of one or more planets is the observed outcome. 

\subsection{Dependence on the inclination of the binary}
 
\begin{figure}
\centering
\includegraphics[width=\columnwidth]{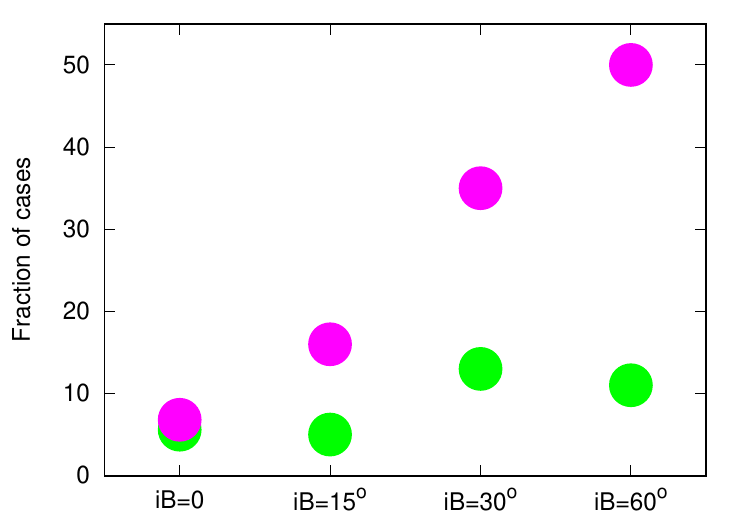}
\caption{The green filled circles mark the fraction of cases in which the inner planet is circularized to a very close orbit (Hot/Warm Jupiter) either during the integration (dynamical tides) or on a longer timescale (Kozai migration). The magenta filled circles show the fraction of cases where the planet is in a Kozai state with the minimum pericenter within 0.5 au.}
\label{fig:incli}
\end{figure}

Figure~\ref{fig:incli} illustrates the cases where 1) the inner planet is circularized to a close orbit and 2) bodies in a Kozai state with a close pericenter. The circularization into a Hot/Warm Jupiter state may happen on a short timescale via dynamical tides or on a longer timescale thanks to a Kozai state with the binary star having a pericenter distance smaller than 0.05 au.  We consider {four different values of the binary inclination i.e., $i_B=0^{\circ}$, $i_B=15^{\circ}$, $i_B=30^{\circ}$ and $i_B=60^{\circ}$. The higher values of inclination may be problematic for the formation of giant planets around the primary star but we will not investigate the issue in this paper and focus only on the dynamical aspects.}  


According to Fig.\ref{fig:incli}, there is a significant increase in the number of tidally circularized planets only when {$i_B \geq 30^{\circ}$ while the number of cases in a Kozai state with pericenter $q$ lower than 0.5 au  steadily increases with the binary inclination. When $i_B \leq 30^{\circ}$ the planets which end up in a Kozai state with the companion star and with $q \leq 0.5$ au are those which, after planet--planet scattering, have their inclination excited beyond the critical value $i_B \sim 39.2^{\circ}$. For higher values of $i_B$ the opposite occurs and some planets end up in non-Kozai states (or with $q \geq 0.5$ au).   }

\section{Initial random distribution of the semi--major axis of the inner planet \texorpdfstring{\lowercase{$a_1$}}{R}}

The initial value of $a_1$ for which a system of three planets during its evolutionary history becomes unstable and undergoes close encounters is difficult to be predicted. It depends on the
formation site of the cores and their growth tracks (see examples in  \cite{bitsch2015,mordasini2018, Pirani2019}) and on the migration by interaction with the  circumstellar disk. Planet--Planet scattering  may occur when the planets are still migrating within the disk  \citep{marzari2010} or when the disk dissipates after the planets have reached a stationary but dynamically unstable state.

In binaries, both planet formation and migration may be significantly different respect to the single star cases. Planetesimal formation, pebble and planetesimal accumulation into cores and gas accretion can be significantly affected by the gravity of the secondary star (see for example \cite{marzari2000, thebault2006, paard2008, kley2008,  thebault2006, marzari2012,  picogna2013, thebault2015, galax2019})  as well as migration by interaction with the disk. In particular, the secondary star truncates the circum--primary disk and reduce the available dynamical range in semi--major axis where planets can form, evolve and possibly migrate without becoming immediately unstable. In this scenario, it is  difficult to derive reasonable initial conditions for systems of three planets before they undergo P--P scattering and even adopt constraints from the observed present distribution around single stars. To perform a realistic modeling of the final orbital distribution of planetary systems undergoing P--P scattering in binaries, it may be reasonable to adopt an initial distribution which depends on the extent of disk truncation induced by the companion star. In this view, we have run additional cases where we select randomly the initial value of $R$  between  0.005 and and 0.025 (e.g. when $a_B=200$, $a_1$ is randomly chosen between 1 and 5 au). This is an arbitrary choice but it can be considered a first step towards a better prediction of a potential trend in observations. Deviations from the distribution we obtain with a random initial sampling within the stability limit must be ascribed to peculiarities in the migration and formation processes. 

\begin{figure}
\centering
\includegraphics[width=\columnwidth]{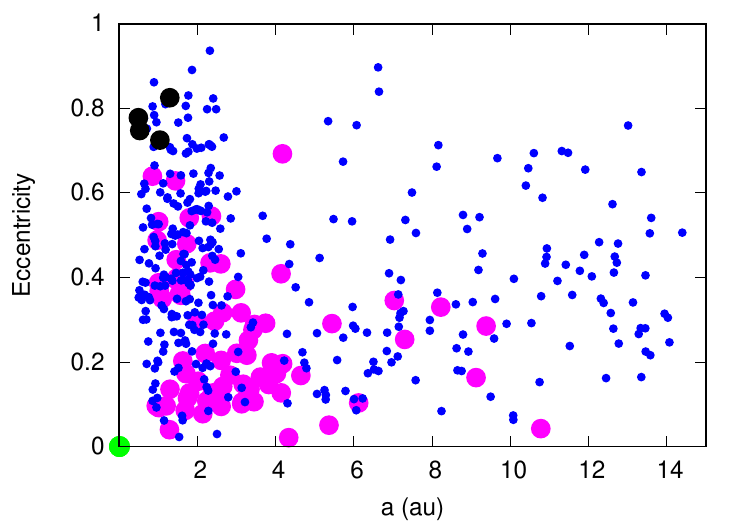}
\caption{Surviving planets in the $[a,e]$- plane in the case with $a_B=200$ au and $e_B = 0.4$. The magenta filled circles are the outcome of collisions, the blues filled circles are the outcome of scattering, the filled black circles are Kozai cases with pericenter smaller than 0.05 au, the green circles are planets which are quickly circularized by dynamical tides.}
\label{fig:randoma200}
\end{figure}

In Fig.\ref{fig:randoma200} we show the final orbital distribution in the $[a,e]$-plane for the case  with $a_B = 200$ au and $e_B=0.4$. In different colors and size of the filled circles are shown: 1) planets which are outcome of P--P scattering with ejection of one or two initial bodies (blue filled circles), 2) more massive planets consequence of a collision and subsequent merging (magenta filled circles), and 3) tidally circularized planets (green filled circles, all concentrated at $a = 0.02$ au and zero eccentricity), and 4)  planets in Kozai resonance with the binary and pericenter less than 0.05 au 
(black filled circles) which may be circularized on a longer timescale. 

To test more quantitatively the final orbital distribution, we have computed histograms of the semi--major axis, eccentricity and inclination of the planets for four different initial configurations which, according to the R--scaling described in Section~3, have the same dynamical characteristics in terms of instability onset. 

\begin{figure*}

\begin{tabular}{cc}
\hspace{0cm}
\includegraphics[angle=0,width=\columnwidth]{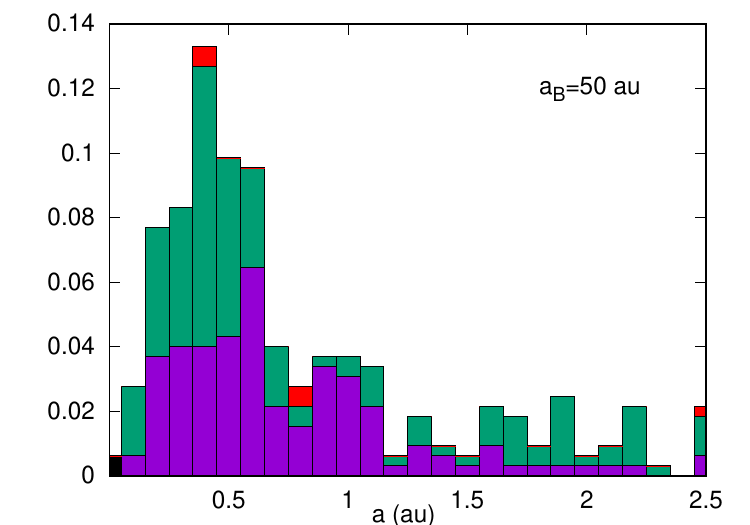} & \hspace{-0.5cm}  \includegraphics[angle=0,width=\columnwidth]{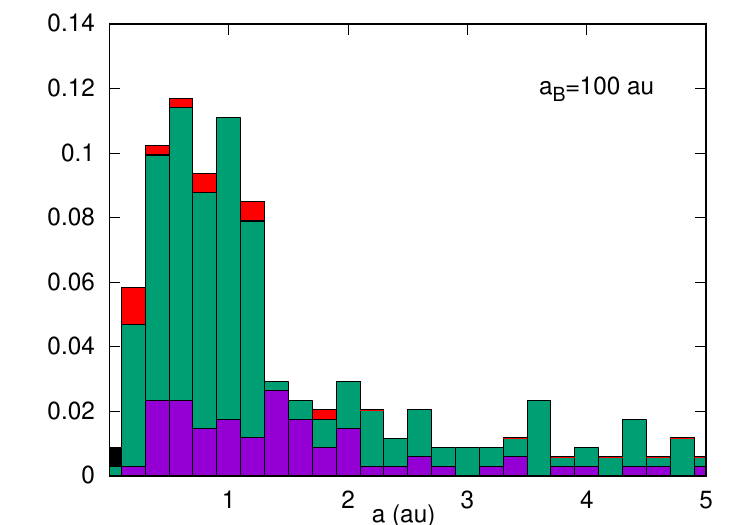} \\  
(a) & (b) \\[6pt] 
\hspace{0cm}
\includegraphics[angle=0,width=\columnwidth]{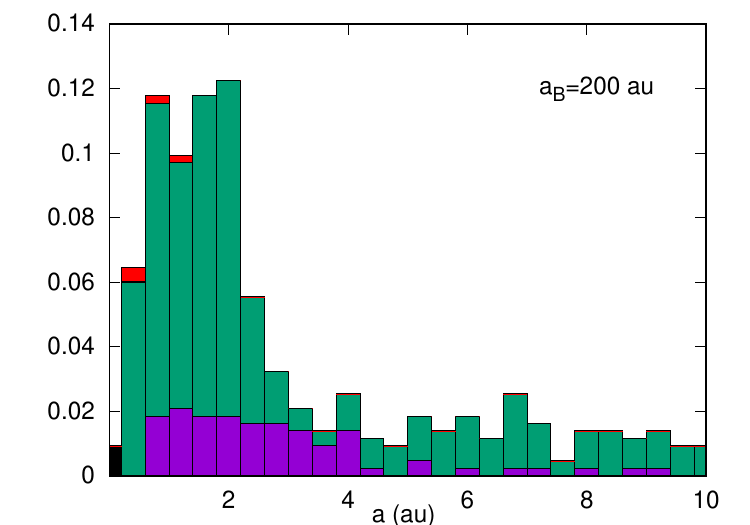} & \hspace{-0.5cm}
 \includegraphics[angle=0,width=\columnwidth]{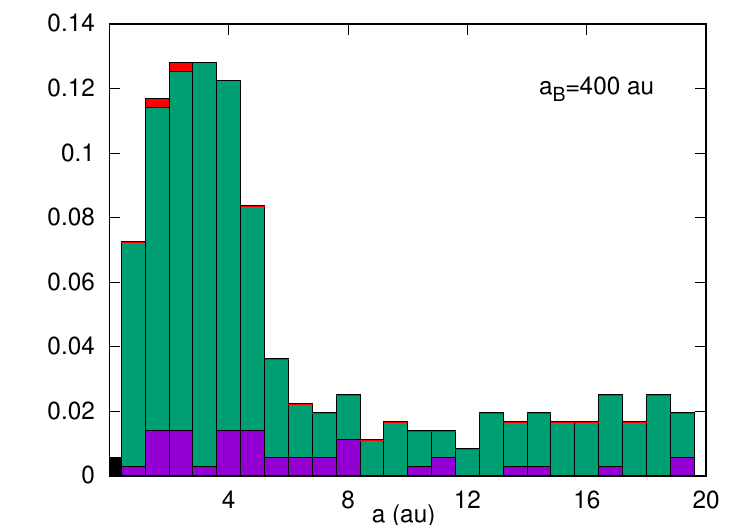} \\
(c) & (d)  \\[6pt]
\end{tabular}
\caption{Normalized histograms showing the final distribution of the semi--major axis of the surviving planets for the case with $a_B=50,100,200,400$ au, from top to bottom. The magenta bars give the fraction of cases where a collision occurs, the green bars are pure P--P scattering, the black bars indicate the planets which are circularized by tides and the red bars are the Kozai cases which may be circularized over a long timescale. }
\label{fig:histoa}
\end{figure*}

In Fig.\ref{fig:histoa} we show the histograms of the final distribution of the semi--major axes of the surviving planets in the four different cases with $a_B = 50,100,200,400$ au and $e_B=0.4$. The random initial sampling of $R$ leads, at the end of the chaotic phase, to a broad peak of scattered planets within $R \sim 0.015$ in all cases. There are differences concerning the fraction of collisions which, as expected, grows significantly for smaller $a_B$ (and then $a_1$ when $R$ is kept constant) and in the number of Kozai cases which may be circularized by tides. However, the overall distribution appears similar suggesting that the scaling with $R$  is approximately valid also for the final distribution of the semi--major axes of the surviving planet.

\begin{figure}
\centering
\includegraphics[width=\columnwidth]{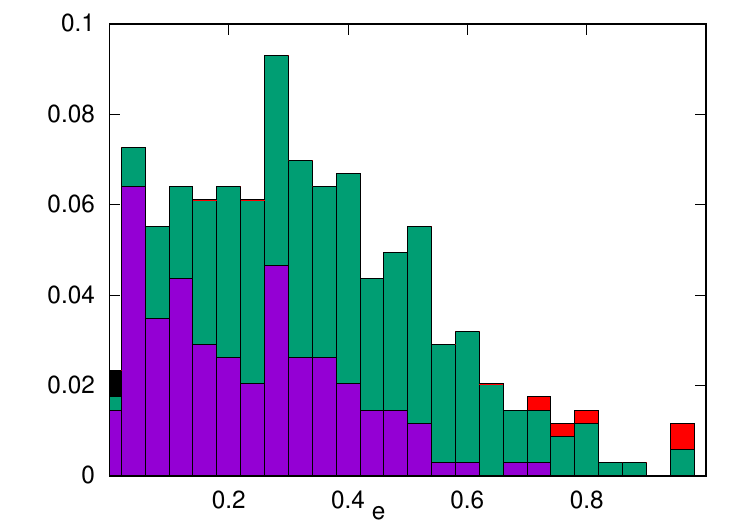}
\includegraphics[width=\columnwidth]{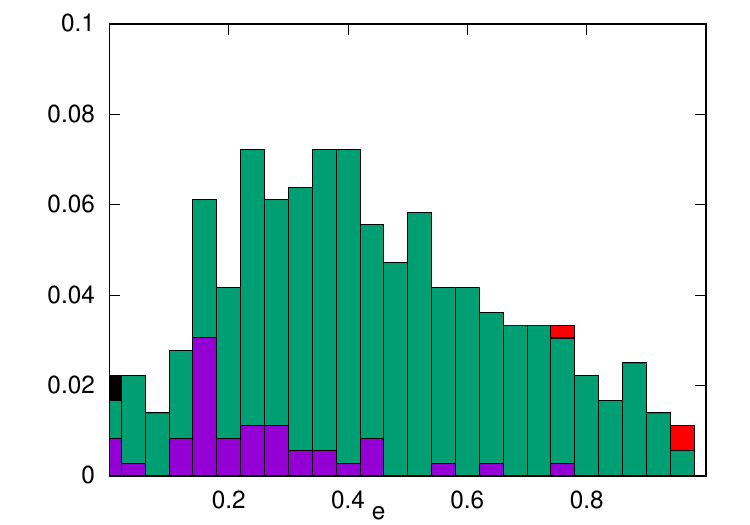}
\caption{Histograms showing the final distribution of the eccentricity of the surviving planets for the case with $a_B=50$ (top panel) and $a_B=400$ au (bottom panel). The meaning of the color bars is the same as in Fig.\ref{fig:histoa}. }
\label{fig:histoe}
\end{figure}

Different is the scenario for the eccentricity distribution. In this case, for small separations like $a_B=50$ au, the high number of planetary collisions shift the eccentricity distribution towards lower values.  The conservation of momentum during a collision leads to lower eccentricities for the merged planets compared to those of the impacting bodies. This is illustrated in  Fig.\ref{fig:histoe} where we compare the case with $a_B=50$ au to that with $a_B=400$ au. This trend may appear from future observations since, according to our modeling, around binaries with small separations a higher number of more massive and less eccentric planets is expected compared to binaries with larger separation. This behaviour  partly invalidate the scaling with $R$ but the effect is noticeable only for very small separations like $a_B=50$ au or smaller. Already for $a_B=100$ (and larger) the eccentricity distributions are very similar since the number of collisions significantly drops. 

\begin{figure}
\centering
\includegraphics[width=\columnwidth]{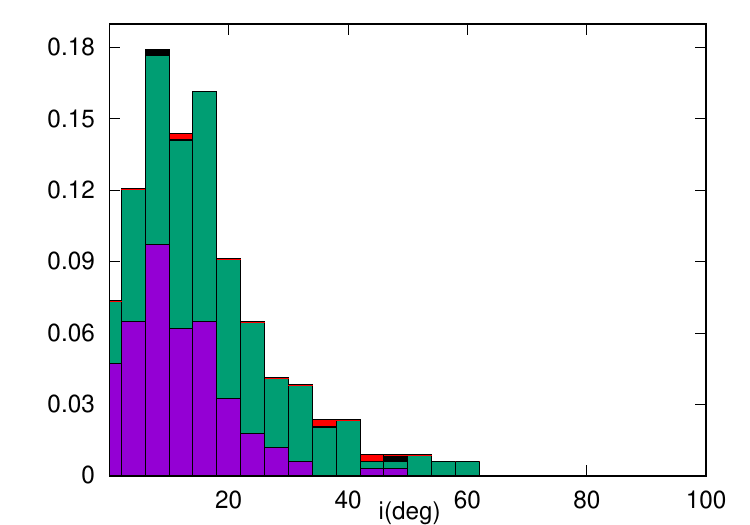}
\includegraphics[width=\columnwidth]{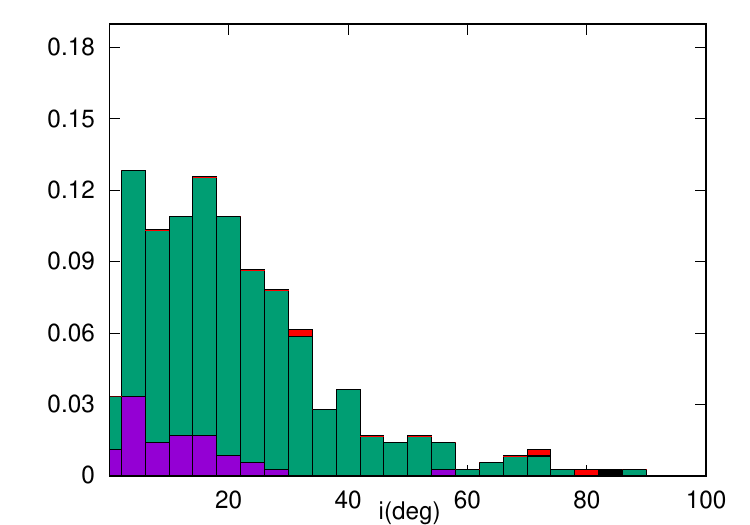}
\caption{Histograms showing the final distribution of the inclination of the surviving planets for the case with $a_B=50$ (top panel) and $a_B=400$ au (bottom panel). The meaning of the color bars is the same as in Fig.\ref{fig:histoa}. }
\label{fig:histoi}
\end{figure}

A similar effect is observed for the inclination distribution shown in  Fig.\ref{fig:histoi}. The case with $a_B=50$ au has a more narrow distribution and there are less cases with high inclination compared to the cases with larger $a_B$. Even in this case it is due to the significant number of collisions occurring during the chaotic phase. 

We can conclude that the semi--major axis distribution approximately scales with $R$ but the eccentricity and inclination distributions for close binaries ($a_B \leq 50$ au) are slightly shifted towards smaller values due to the large number of collisions occurring during the chaotic phase. The fraction of Kozai cases, which should increase by reducing $a_B$ do not really grow because of the increasing number of cases ending up with collisions.

\section{Conclusions}

We have analysed the dynamics and evolution of systems of three planets in binary star systems with particular emphasis on Jupiter--size planets. The first interesting outcome is that the dynamics of the system, in particular its long term stability, can be scaled with the ratio between the inner planet semi--major axis $a_1$ and the binary semi--major axis $a_B$, ratio that we termed $R$. This means that the dynamical behaviour of a given binary system can be extended to any other binary system with different $a_B$ as long as the semi--major axis of the inner planet is changed in order to keep $R$ constant and the initial semi-major axes of the other two planets are scaled accordingly. Systems with the same value of $R$ have instability times, before it occurs the first close encounter, which grows with the Keplerian period for increasing $a_B$, while the stable regions in the phase space are superimposable. This result can be applied to any group of three planets with similar mass, whether they are Jupiter, Neptune or super--Earth size planets.    The $R$ scaling has the advantage of reducing the number of degrees of freedom when studying the evolution of the planets in binary systems and their stability. 

When it comes to P--P scattering, the planets during their evolution may move very close to the star and additional physics comes into play like tides and general relativity. The effects of both depend  on the values of the semi--major axis of the planets, in particular the inner one. The companion star also significantly affects the chaotic evolution of the planets by draining some of the planet orbital energy via close encounters and leading to a broader final distribution of the semi--major axis of the inner surviving planet. In addition, the encounters with the star favors the ejection of the planets and less two--planet systems are expected in binary systems compared to single stars. 

To be predictive of the final orbital distribution of planets in real systems, we have performed, for any given binary configuration, a large number of numerical simulations with random initial semi--major axis of the inner planet.
By comparing the orbital distribution of the surviving planets after the chaotic phase we find that there is not a significant difference in the final semi--major axis distribution among the cases with the same value of the scaling parameter $R$. However, the eccentricity and inclination distributions for small values of $a_B$ are shifted towards smaller values due to the increasing number of collisions which, after merging of the two impactors, lead to lower values of eccentricity and inclination. However, the $R$ scaling is approximately reinstated when  $a_B \geq 100$ au since collisions are less frequent and the dynamics is less affected by them. 

The simulations on P--P scattering are focused on Jupiter---size planets and, according to the properties of HD 41004 A and Gliese 86, there seems to be enough mass in solids to build multiple massive planetary systems even from disks around close binaries.  It would be interesting to investigate the evolution of less massive multiple systems made of Neptune--size planets and super--Earths. These configurations should follow the $R$ scaling even at small binary separations since tides are expected to be significantly weaker and collisions less frequent due to the reduction of the cross section. However, this should be investigated.

\section*{Acknowledgements}
We thank the referee Sean Raymond whose comments and suggestions helped to improve the paper. 
K.G. thanks the Centre of Informatics Tricity Academic Supercomputer \& Network (Gda\'nsk, Poland) for a computing grant and resources.
N.M. was supported by JSPS KAKENHI Grant Number JP21K03650 and JP18H05438. 
We acknowledge the use of the REBOUND code by Hanno Rein et al. to plot Fig.~\ref{fig:geometry}.
\section*{Data Availability}

The data underlying the research results described in the article will 
be shared upon reasonable request to the authors.

\bibliographystyle{mnras}
\bibliography{references}
\label{lastpage}
\end{document}